\documentclass[11pt]{elsarticle}
\usepackage{amsfonts,amssymb}
\usepackage{amsmath,amscd}
\usepackage{amsmath}
 \newtheorem{thm}{Theorem}
 \newtheorem{cor}{Corollary}

\usepackage{graphicx}
\usepackage{epstopdf}
\usepackage{subfigure}
\usepackage{mathrsfs}

\newcommand\dif{\mathrm{d}}

\makeatletter

\begin{document}
\begin{frontmatter}

\title{The network level reproduction number for infectious diseases with both vertical and horizontal transmission }

\author {Ling Xue\corref{a}}
 \ead{lxue@ksu.edu}
\author {Caterina Scoglio}
 \cortext[a]{Corresponding Author}

 \ead{caterina@k-state.edu}
 \address {Department of Electrical $\&$  Computer Engineering,\\ Kansas State University, \ U.S. \ 66506}

\date{}
\begin{abstract}
A wide range of infectious diseases are both vertically and horizontally transmitted.   Such diseases are spatially transmitted via multiple species in heterogeneous environments, typically described by  complex meta-population models.  The reproduction number, $R_0$, is a critical metric predicting whether the disease can invade the meta-population system.  This paper presents the reproduction number for a generic disease vertically and horizontally transmitted among multiple species in heterogeneous networks, where nodes are locations, and links reflect outgoing or incoming movement flows.  The metapopulation model for vertically and horizontally transmitted diseases is gradually formulated from  two species,  two-node network models.    We derived  an explicit expression of $R_0$, which is the spectral radius of a  matrix reduced in size with respect to  the original next generation matrix.    The reproduction number is shown to be a function of  vertical and horizontal transmission parameters, and the lower bound is the reproduction number for  horizontal transmission.  As an application, the reproduction number and its bounds for the Rift Valley fever zoonosis, where livestock, mosquitoes, and humans are the involved species are derived.  By computing the reproduction number for different scenarios through numerical simulations, we found the reproduction number  is affected by livestock movement rates  only when parameters are heterogeneous across nodes.  To summarize, our study contributes the reproduction number for vertically and horizontally transmitted diseases in heterogeneous networks.  This explicit expression is easily adaptable to specific infectious diseases, affording insights into disease evolution.

\end{abstract}


\end{frontmatter}
\section{Introduction}
Communicable diseases are readily transmitted from one region to another \cite{Salmani2006, Wang2004}. Population travel continues to influence the temporal and spatial spread of infectious diseases  \cite{Salmani2006, Arino2005}.  Observation of the introduction of infectious agents resulting in spatial spreading of effective infections in different locations at different times \cite{Arino2005}, revealed great economic losses, many animal and human cases, and deaths. Noteworthy examples include the fourteenth century  plague  in Europe \cite{ Salmani2006, Stothers2000} and the sixteenth century smallpox epidemic in the New World  \cite{Salmani2006}. More recent epidemics, including HIV/AIDS and West Nile virus in North America \cite{Petersen2001}, and SARS in Asia \cite{WHO2003}, show infections spreading over vast regions and even jumping continents \cite{Arino2007}.

Many communicable diseases are propagated by two distinct mechanisms: vertical and horizontal transmission  \cite{Busenberg1982}.   Vertical transmission occurs when infection is passed from mother to a portion of offspring \cite{Busenberg1982, El-Doma1999}, often transmitted by insect eggs and/or plant seeds  \cite{Li2001}.  A variety of diseases are transmitted vertically and horizontally, including the human diseases rubella, hepatitis B, Chagas’ disease, and AIDS  \cite{Li2001, BUSENBERG1993}. Vertical transmission is a proven factor in the size and persistence of Rift Valley fever (RVF) epidemic \cite{Chitnis2013}. The prevalence of vertical transmission establishes it as a crucial biological mechanism \cite{BUSENBERG1993}, potentially affecting infectious spreading in elaborate ways  \cite{Anderson1981}.   Therefore, vertical transmission acts to maintain the spread of infection \cite{Anderson1981, Busenberg1988}.
The logical complement of vertical transmission is horizontal transmission. For animal and human diseases  \cite{Li2001}, horizontal transmission is often through direct or indirect contact  with infectious hosts or infectious vectors, such as biting insects  \cite{Li2001}.

Spatially structured models, such as meta-population models or multiple-patch models are widely used in epidemiological modeling to capture the effect of space  \cite{Rohani1999}. Meta-population models describe systems containing spatially discrete sub-populations connected by the movement of individuals between a set of patches or nodes \cite{Gotelli1999, Taylo2012}. Modeling the dynamics of large metapopulations is complex, presenting challenges during analysis \cite{Arino2003}. One approach considers the mobility of individuals between discrete regions  \cite{Arino2003}, creating a directed network where nodes represent locations and links are movements between locale \cite{Arino2003}.   The importance of tracking mobility rates and movement patterns is highlighted in the foot-and-mouth outbreak of $2001$ in the United Kingdom \cite{Arino2007}.  There, infected cattle were widely distributed before the movement ban was announced \cite{Keeling2001}, prompting the necessary development of a transportation network capturing the spatial spread of foot-and-mouth disease \cite{Arino2007}.

Numerical tools are widely used to obtain quantitative results and analytic tools are used to  understand model behaviors   \cite{Arino2005}. The reproduction number, $R_0$, defined as the average number of new infected individuals produced by one infectious individual, in a population with only susceptibles \cite{Diekmann2010},  is  arguably the most important quantity in communicable disease modeling \cite{Diekmann2010}. Theoretically, $R_0$ plays an important role in analyzing the dynamics of an epidemic  \cite{Diekmann2010}. It is a quantity commonly used to estimate the dynamics of emerging infectious diseases at the beginning of an outbreak, aiding in the design of control strategies for established infections \cite{Diekmann2010}. The next generation method developed by \cite{diekmann1990definition},  ~\cite[Chapter 5]{diekmann2000mathematical}  and popularized by \cite{van2002reproduction} is one of many methods applied to compute the reproduction number  for compartmental models.   This method manages matrix size by including only infected and asymptomatically infected states  \cite{Li2011}.  The next generation matrix relates the number of new cases in various generations and  provides the basis of defining and computing the reproduction number \cite{Diekmann2010}. 

The very little work on the reproduction number for meta-populations with vertical transmission we encountered included the modeling of horizontal and vertical transmission dynamics of a parasite with two differential equations  \cite{Lipsitch1995}. In this special case, the reproduction number is the sum of the reproduction numbers for both types of transmission,  and does not hold for a more complicated situation, such as in the model   \cite{LingXue2010},  where the next generation matrices for the two types of transmission are not both scalars.  As far as we know, an insightful explicit expression of $R_0$ for multiple species meta-population model with complex transmission has not yet been presented.

This paper presents the computation of the reproduction number and its bounds for compartmental models considering diseases with complex transmission.    We consider meta-populations consisting of discrete, well-mixed subpopulations.  We assume that individuals move between different nodes and the disease can be transmitted within a node.  An $n-$node compartmental model incorporates $h$  species, of which  $g$ species transmit a disease both vertically and horizontally and other  species only transmit horizontally.   All sojourn times are taken to be exponentially distributed, and vertical transmission is restricted to the egg stage with exponential duration. Presented here is  a general network-level model applicable when studying the temporal-spatial propagation of an infectious disease with multi-species, vertical and horizontal transmission, where the reproduction number is derived as a function of the two types of transmission parameters.  Finally, the exact value and bounds of the reproduction number for the RVF meta-population model are computed and factors affecting the reproduction number are analyzed.  We found the upper bound depends on both horizontal and vertical transmission, while the lower bound is determined solely by horizontal transmission.

The contribution of our work is summarized as follows:
\begin{enumerate}
\item An explicit expression of the reproduction number considering vertical and horizontal transmission in a general multi-species, meta-population model is derived.
\item This formula for the reproduction number is applied to an RVF meta-population model to compute $R_0$ and its bounds.
\item Numerical simulations  show that livestock movement rates only affect $R_0$ for heterogeneous networks relative to  disease parameters.
\end{enumerate}

Our work facilitates computation of the exact reproduction number in a meta-population model with complex disease transmission.

The paper is organized as follows.  Section $ \ref{section:model}$ describes the next generation matrix approach used to derive an explicit expression of  the reproduction number, and presents the general meta-population model beginning with two species, two-node network models,  as well as computing   the reproduction number.  In Section $\ref{section:RVFapplication}$, we apply our $R_0$ formula to the RVF meta-population model, computing $R_0$ and its bounds.  The effects of livestock movement, heterogeneities of parameters, and the size of a network on the reproduction number are also studied through simulations.  Section $ \ref{section:result}$ provides a  summary and discussion of mathematical derivations and simulation results.

\section{The reproduction  number for diseases with both vertical and horizontal transmission}
\label{section:model}

One  frequently used method computes the reproduction number as the spectral radius  of the next generation matrix  \cite[Chapter 5]{diekmann2000mathematical}, \cite{heffernan2005perspectives, Diekmann2010}.  For the ease of computation, only the compartments corresponding
to infected and asymptomatically infected compartments are considered \cite{Diekmann2010}.   First, the original nonlinear ODE  system is decomposed into two column vectors $\mathscr{F}=(\mathscr{F}_i)$ and $\mathscr{V}=(\mathscr{V}_i)$,
where  $\mathscr{F}_i $ is the $ i^{th}$ row of $\mathscr{F}$ representing the rate at which  new infections appear in compartment $i$, and $\mathscr{V}_i$   is the $ i^{th}$ row of $\mathscr{V}$. Moreover,  $\mathscr{V}_i=\mathscr{V}_i^{-}-\mathscr{V}_i^{+}$, where $\mathscr{V}_i^{-}$ represents the rate at which individuals  transfer  out of compartment $i$, and $\mathscr{V}_i^{+}$ is the rate at which individuals transfer into compartment $i$ \cite{van2002reproduction}.  Assume that the number of infected and asymptomatically infected compartments is $m$. The 
Jacobian matrices $F$ denoting transmission, and $V$ denoting transition  \cite{Diekmann2010} are defined as:
\begin{equation}\label{Jacobian}
F= [\frac{\partial \mathscr{F}_i (x^0)}{\partial x_j}], \quad V= [\frac{\partial \mathscr{V}_i(x^0)}{\partial x_j}],
\end{equation}
where $x^0$ represents the disease free equilibrium (DFE), and $x_j$ is the number or proportion of infected individuals in compartment $j$, $j=1, 2, \cdots, m$.

The spectral radius of a matrix $A$ is denoted by $\rho(A)$. The reproduction number, $R_0$, is defined as  $\rho(FV^{-1})$ \cite{diekmann1990definition}. To understand entries of  $FV^{-1}$, called the next generation matrix, consider the consequence of an infected individual introduced into compartment $k$ in a population at DFE \cite{van2002reproduction}.  The $(i, j)$ entry of $F$ represents the   rate at which  new  infected individuals in  compartment $i$ are produced by infected individuals in compartment $j$ \cite{van2002reproduction}. The $(j, k)$ entry of $V^{-1}$ represents the average time that an  infected  individual stays in compartment $j$ \cite{van2002reproduction}. Hence, the  $(i, k)$ entry of $FV^{-1}$ represents the expected number of new infections  in compartment $i$ resulting from the infected individual originally introduced  into compartment $k$ \cite{van2002reproduction}, where $i, k=1, 2, \cdots, m$.  Note that matrix $F$  is nonnegative and $V$ is proved to be a  nonsingular M-matrix \cite{van2002reproduction}. Recall that an $n \times n$ matrix  $A$  is an M-matrix if it  can be expressed in the form  $A=sI-B$, such that matrix $B$ is  non-negative, and $s  \geqslant  \rho(B)$ \cite{Plemmons1977}.

Next, we illustrate computational procedures for finding  $R_0$ using the next generation matrix method for susceptible-exposed-infectious-recovered (SEIR) compartmental models, assuming a disease is transmittable within a species and between different species, and movement rates for all species are independent of disease status.   Daily time steps are used in all models.

\subsection{Models for two species in two nodes}
\label{subsection:two species}
We present two applications of a simplified system for a disease involving two species in a two-node network with movement between the two nodes.  In the first example, $R_0$ is computed while assuming  only horizontal transmission is taking place.  In the second example,  the first  model is extended by introducing vertical transmission into one species.  The reproduction number  is  then computed.

\subsubsection{$R_0$ for two species with only horizontal transmission }
\label{subsubsection:no vertical transmission}
Below,  a compartmental model for an infectious disease incorporating  four compartments $(J=S, E, I, R)$, two species $( k=1, 2)$,  two nodes $( i=1, 2 )$, and only horizontal transmission is presented.   The differential equations representing the dynamic behavior are:
\allowdisplaybreaks
\begin{align}
 \frac{dS_{ki}}{dt}&=r_{ki}-\beta_{1ki}S_{ki}I_{1i}/N_{1i}-\beta_{2ki}S_{ki}I_{2i}/N_{2i}
-d_{ki}S_{ki}+\sum^2_{j=1, j \neq i}\omega_{kji}S_{kj}-\sum^2_{j=1, j \neq i}\omega_{kij}S_{ki} \label{equation:susceptible}\\
 \frac{dE_{ki}}{dt}&=\beta_{1ki}S_{ki}I_{1i}/N_{1i}+\beta_{2ki}S_{ki}I_{2i}/N_{2i}
-\varepsilon_{ki}E_{ki}-d_{ki}E_{ki}+\sum^2_{j=1, j \neq i}\omega_{kji}E_{kj}-\sum^2_{j=1, j \neq i}\omega_{kij}E_{ki}\\
  \frac{\dif I_{ki}}{\dif t} &=\varepsilon_{ki}E_{ki}-\gamma_{ki}I_{ki}-d_{ki}I_{ki}+\sum^2_{j=1, j \neq i}\omega_{kji}I_{kj}-\sum^2_{j=1, j \neq i}\omega_{kij}I_{ki}\\
\frac{\dif R_{ki}}{\dif t} &=\gamma_{ki}I_{ki}-d_{ki}R_{ki}+\sum^2_{j=1, j \neq i}\omega_{kji}R_{kj}-\sum^2_{j=1, j \neq i}\omega_{kij}R_{ki}. \label{equation:recovered}
\end{align}

The number of  newborn individuals of species $k$ in node $i$ per day  is denoted by $r_{ki}$.  The number of  species $k$ individuals  in node $i$ of  compartment $J$  is denoted by $J_{ki}$,  and
the total number  of species $k$ individuals  in  node $i$ is  $N_{ki}=S_{ki}+E_{ki}+I_{ki}+R_{ki}$.  Total individuals  of species $k$  infected daily  in node $i$ by species $1$ and species $2$  are $ \beta_{1ki}S_{ki}I_{1i}/N_{1i}$ and $\beta_{2ki}S_{ki}I_{2i}/N_{2i}$, respectively. The  number of deaths from each compartment $J$  per day is $d_{ki}J_{ki}$. After the  incubation period, $\varepsilon_{ki} E_{ki}$  individuals transfer to infected compartment  daily. Following the  infection period, $\gamma_{ki} I_{ki}$  recover from the infection each day. Movement rates for species $k$ individuals in compartment $J$    in and out of node $i$ are  $\sum^2_{j=1, j \neq i}\omega_{kji}J_{kj}$ and $\sum^2_{j=1, j \neq i}\omega_{kij}J_{ki}$, respectively.

Species $k$ quantity  in compartment $J$ and  the total number in node $i$ at DFE are denoted by $J_{ki}^0$ and  $N_{ki}^0$, respectively. To compute $R_0$ using the next generation matrix method, we need to  prove the existence   and uniqueness of DFE.
At DFE,
$ S^0_{1i}=N^0_{1i}$, and
$S^0_{2i}  =N^0_{2i}$, as $E^0_{1i}=I^0_{1i}=R^0_{1i}=E^0_{2i}=I^0_{2i}=R^0_{2i}=0$.
This  is a special case of the model  for  Theorem \ref{theorem:N2} (see appendix), which determines the existence of    a   unique solution  $[ N^0_{1i}   \ \ \  N^0_{2i}]^T$.

The equations related to exposed and infected compartments are ordered:
\begin{equation*}
\frac{d}{dt}\left[
\begin{array}{rllllllllllllllllllllllllllllllllllllllll}
E_{11} &E_{12} &E_{21} &E_{22}&I_{11} & I_{12 } & I_{21} & I_{22}
\end{array}\right]^T
= \mathscr{F}_H - \mathscr{V}_H,\ {\rm where}
\end{equation*}
\begin{equation*}
\mathscr{F}_H=\begin{bmatrix}
\beta_{211}S_{11}I_{21}/N_{21}+\beta_{111}S_{11}I_{11}/N_{11} \\
\beta_{212}S_{12}I_{22}/N_{22}+\beta_{112}S_{12}I_{12}/N_{12} \\
\beta_{121}S_{21}I_{11}/N_{11}+\beta_{221}S_{21}I_{21}/N_{21} \\
\beta_{122}S_{22}I_{12}/N_{12}+\beta_{222}S_{22}I_{22}/N_{22} \\
0 \\
0 \\
0 \\
0 \\
\end{bmatrix},\quad
\mathscr{V}_H =\begin{bmatrix}
d_{11}E_{11}+\varepsilon_{11}E_{11}+\omega_{112}E_{11}-\omega_{121}E_{12}\\
d_{12}E_{12}+\varepsilon_{12}E_{12}+\omega_{121} E_{12}-\omega_{112} E_{11}\\
d_{21}E_{21}+\varepsilon_{21}E_{21}+\omega_{212} E_{21}-\omega_{221} E_{22} \\
d_{22}E_{22}+\varepsilon_{22}E_{22}+\omega_{221} E_{22}-\omega_{212} E_{21}\\
-\varepsilon_{11}E_{11}+ d_{11}I_{11}+\gamma_{11}I_{11}+\omega_{112}I_{11}-\omega_{121} I_{12}\\
-\varepsilon_{12}E_{12}+d_{12}I_{12}+\gamma_{12}I_{12}+\omega_{121}I_{12}-\omega_{112} I_{11}\\
-\varepsilon_{21}E_{21} + d_{21}I_{21}+\gamma_{21}I_{21}+\omega_{212} I_{21}-\omega_{221} I_{22} \\
-\varepsilon_{22}E_{22} + d_{22}I_{22}+\gamma_{22}I_{22}+\omega_{221} I_{22}-\omega_{212} I_{21}\\
\end{bmatrix}.
\end{equation*}
By (\ref{Jacobian}), the Jacobian matrices for this model are:
\begin{equation}\label{FHVH}
F_H=\begin{bmatrix}
  0_{4\times 4}& \mathcal{A}\\
0&0_{4\times 4}\\
\end{bmatrix},\quad V_H=\begin{bmatrix}
 \oplus _{k=1}^{2}M_k & 0\\
-  \oplus _{k=1}^{2}(\oplus _{i=1}^{2}\varepsilon_{ki}) & \oplus _{k=1}^{2}X_k
\end{bmatrix},
\end{equation}
where the  symbol  $ \bigoplus$ represents the  direct sum of matrices, i.e., $ A\bigoplus B$ =$\left[ \begin{array}{cccccccccccccccc}
A&0  \\
0&B\\
\end{array} \right]$ for matrices $A$ and $B$.  The subscript of the zero blocks, $4 \times 4$,  indicates the size of the block. Matrices $  \mathcal{A}$, $M_k$ and $X_k$ are:
\begin{equation*}
  \mathcal{A}=\begin{bmatrix}
    \beta_{111}\frac{S_{11}^0}{N_{11}^0}&0&\beta_{211}\frac{S_{11}^0}{N_{21}^0}&0\\
0&\beta_{112}\frac{S_{12}^0}{N_{12}^0}&0&\beta_{212}\frac{S_{12}^0}{N_{22}^0}\\
\beta_{121}\frac{S_{21}^0}{N_{11}^0}&0&\beta_{221}\frac{S_{21}^0}{N_{21}^0}&0\\
0&\beta_{122}\frac{S_{22}^0}{N_{12}^0}&0&\beta_{222}\frac{S_{22}^0}{N_{22}^0}\\
  \end{bmatrix},
\end{equation*}

 \begin{equation}\label{M1M2}
M_1= \left[ {%
\begin{array}{cccccccccccccccc}
d_{11}+\varepsilon_{11}+\omega_{112}&-\omega_{121} \\
-\omega_{112} & d_{12}+\varepsilon_{12}+\omega_{121}\\
\end{array}
} \right] , \ \  M_2= \left[ {%
\begin{array}{cccccccccccccccc}
d_{21}+\varepsilon_{21}+\omega_{212} & -\omega_{221}  \\
-\omega_{212}  & d_{22}+\varepsilon_{22}+\omega_{221} \\
\end{array}
} \right] ,
\end{equation}
\begin{equation}\label{X1X2}
X_1= \left[ {%
\begin{array}{cccccccccccccccc}
d_{11}+\gamma_{11}+\omega_{112}&-\omega_{121}    \\
-\omega_{112}  &d_{12}+\gamma_{12}+\omega_{121}\\
\end{array}
} \right],  \ \  X_2= \left[ {%
\begin{array}{cccccccccccccccc}
d_{21}+\gamma_{21}+\omega_{212}& -\omega_{221} \\
  -\omega_{212} &d_{22}+\gamma_{22}+\omega_{221}\\
\end{array}
} \right].
\end{equation}
Because the  matrices $M_1$, $M_2$, $X_1$, and $X_2$ are all invertible, we can readily check:
\begin{equation*}
V_H^{-1}=\left[ {%
\begin{array}{cccccccccccccccc}
\oplus _{k=1}^{2}M_k^{-1}&0\\
 \oplus_{k=1}^2\mathcal{Z}_k& \oplus  _{k=1}^{2} X_k^{-1}\\
\end{array}
} \right],
\end{equation*}
where $\mathcal{Z}_k=X_k^{-1}(\oplus  _{i=1}^{2}\varepsilon_{ki})M_k^{-1}$. The spectral radius of the next generation matrix $F_HV_H^{-1}$  is:
\begin{eqnarray*}
\rho(F_HV_H^{-1})=\rho(\begin{bmatrix}
  0_{4\times 4}& \mathcal{A}\\
0&0_{4\times 4}\\
  \end{bmatrix}\begin{bmatrix}
\oplus _{k=1}^{2}M_k^{-1}&0\\
 \oplus_{k=1}^2\mathcal{Z}_k& \oplus  _{k=1}^{2} X_k^{-1}\\
\end{bmatrix})
  =\rho(
 \mathcal{A} (\oplus_{k=1}^{2}\mathcal{Z}_k)).
\end{eqnarray*}
Therefore,
\begin{equation}R_0^H:=\rho(F_HV_H^{-1})=\rho(
 \mathcal{A} (\oplus _{k=1}^{2}\mathcal{Z}_k)), \label{R0H2species}
 \end{equation}
  where $R_0^H$  is the reproduction number for horizontal transmission.

\subsubsection{$R_0$ for two species  with vertical transmission in one species }\label{sec2.2.4}
\label{subsubsec:two patches with vertical transmission}

We keep the model for species $2$ (Equation (\ref{equation:susceptible}) to (\ref{equation:recovered}) with $k=2$), while extending the model for species $1$ by incorporating vertical transmission.  The model for species $1$ is:
\allowdisplaybreaks
\begin{align}
 \frac{dP_{1i}}{dt} &=r_{1i} -b_1q_{1i}I_{1i}-\theta_{1i} P_{1i} \label{equation:2.1.2P}\\
 \frac{dQ_{1i}}{dt} &=b_{1i}q_{1i}I_{1i} -\theta_{1i} Q_{1i}\\
 \frac{dS_{1i}}{dt}&=\theta_{1i} P_{1i}-\beta_{11i}S_{1i}I_{1i}/N_{1i}-\beta_{21i}S_{1i}I_{2i}/N_{2i}
-d_{1i}S_{1i}+\sum^2_{j=1, j \neq i}\omega_{1ji}S_{1j}-\sum^2_{j=1, j \neq i}\omega_{1ij}S_{1i}\\
 \frac{dE_{1i}}{dt}&=\beta_{11i}S_{1i}I_{1i}/N_{1i}+\beta_{21i}S_{1i}I_{2i}/N_{2i}
-\varepsilon_{1i}E_{1i}-d_{1i}E_{1i}+\sum^2_{j=1, j \neq i}\omega_{1ji}E_{1j}-\sum^2_{j=1, j \neq i}\omega_{1ij}E_{1i}\\
  \frac{\dif I_{1i}}{\dif t} &=\theta_{1i} Q_{1i}+\varepsilon_{1i}E_{1i}-\gamma_{1i}I_{1i}-d_{1i}I_{1i}+\sum^2_{j=1, j \neq i}\omega_{1ji}I_{1j}-\sum^2_{j=1, j \neq i}\omega_{1ij}I_{1i}\\
\frac{\dif R_{1i}}{\dif t} &=\gamma_{1i}I_{1i}-d_{1i}R_{1i}+\sum^2_{j=1, j \neq i}\omega_{1ji}R_{1j}-\sum^2_{j=1, j \neq i}\omega_{1ij}R_{1i} \label{equation:2.1.2R}.
\end{align}

The number of eggs laid by  species $1$ per day is   denoted as  $r_{1i}$,  including  $b_{1i}q_{1i}I_{1i} $ infected eggs, and  $r_{1i}-b_{1i}q_{1i}I_{1i}$ uninfected eggs.  After the development period,  $\theta_{1i} P_{1i}$ eggs develop into susceptible  adults, and $\theta_{1i}Q_{1i} $ eggs develop into infected  adults daily. The interpretations of other terms are the same as corresponding terms described  in Section  $\ref{subsubsection:no vertical transmission}$.

At DFE, $Q_{1i}^0=E_{1i}^0=I_{1i}^0=R_{1i}^0=E_{2i}^0=I_{2i}^0=R_{2i}^0=0$, $
S^0_{1i}=N^0_{1i}$, and
$S^0_{2i}  =N^0_{2i}$. Since this is another special case of the model  for Theorem \ref{theorem:N2}, a unique solution $[ N^0_{1i}   \ \ \  N^0_{2i}]^T$ exists.
In our second model, the equations related to exposed and infected compartments are ordered:
\begin{equation*}
\frac{d}{dt}\left[
\begin{array}{rllllllllllllllllllllllllllllllllllllllll}
Q_{11} &Q_{12}  & E_{11} &E_{12} &E_{21} &E_{22}&I_{11} & I_{12 } & I_{21} & I_{22}
\end{array}\right]^T
= \mathscr{F} - \mathscr{V},\ {\rm where}
\end{equation*}
\begin{equation*}
\mathscr{F}=\begin{bmatrix}
b_{11}q_{11}I_{11}\\
b_{12}q_{12}I_{12}\\
\beta_{211}S_{11}I_{21}/N_{21}+\beta_{111}S_{11}I_{11}/N_{11} \\
\beta_{212}S_{12}I_{22}/N_{22}+\beta_{112}S_{12}I_{12}/N_{12} \\
\beta_{121}S_{21}I_{11}/N_{11}+\beta_{221}S_{21}I_{21}/N_{21} \\
\beta_{122}S_{22}I_{12}/N_{12}+\beta_{222}S_{22}I_{22}/N_{22} \\
0 \\
0 \\
0 \\
0 \\
\end{bmatrix},\
\mathscr{V} =\begin{bmatrix}
\theta_{11}Q_{11}\\
\theta_{12}Q_{12}\\
d_{11}E_{11}+\varepsilon_{11}E_{11}+\omega_{112}E_{11}-\omega_{121}E_{12}\\
d_{12}E_{12}+\varepsilon_{12}E_{12}+\omega_{121} E_{12}-\omega_{112} E_{11}\\
d_{21}E_{21}+\varepsilon_{21}E_{21}+\omega_{212} E_{21}-\omega_{221} E_{22} \\
d_{22}E_{22}+\varepsilon_{22}E_{22}+\omega_{221} E_{22}-\omega_{212} E_{21}\\
-\theta_{11}Q_{11}-\varepsilon_{11}E_{11}+ d_{11}I_{11}+\gamma_{11}I_{11}+\omega_{112}I_{11}-\omega_{121} I_{12}\\
-\theta_{12}Q_{12}-\varepsilon_{12}E_{12}+d_{12}I_{12}+\gamma_{12}I_{12}+\omega_{121}I_{12}-\omega_{112} I_{11}\\
-\varepsilon_{21}E_{21} + d_{21}I_{21}+\gamma_{21}I_{21}+\omega_{212} I_{21}-\omega_{221} I_{22} \\
-\varepsilon_{22}E_{22} + d_{22}I_{22}+\gamma_{22}I_{22}+\omega_{221} I_{22}-\omega_{212} I_{21}\\
\end{bmatrix}.
\end{equation*}
By (\ref{Jacobian}), the Jacobian matrices for this model are:
\begin{equation*}
F=\begin{bmatrix}
  0_{2\times 2}& U_{2 \times 8}\\
0_{8 \times 2}&F_H\\
\end{bmatrix},\quad  V=\begin{bmatrix}
  \oplus _{i=1}^{2}\theta_{1i}&0_{2 \times 8}\\
 W_{8 \times 2}&V_H
\end{bmatrix}.
\end{equation*}
Here $F_H$ and $V_H$ are the matrices in (\ref{FHVH}) and
\begin{equation*}
U= \left[ {%
\begin{array}{cccccccccccccccc}
0_{2\times 4} &\bigoplus_{i=1}^2 b_{1i}q_{1i} &0_{2\times 2}
\end{array}
} \right],\quad
W= \left[ {%
\begin{array}{cccccccccccccccc}
 0_{4 \times 2} \\
-\bigoplus_{i=1}^2 \theta_{1i} \\
 0_{2 \times 2}
\end{array}
} \right].
\end{equation*}
The matrix $V^{-1}$ and the next generation matrix $FV^{-1}$ are:
\begin{equation*}
V^{-1}= \left[ {%
\begin{array}{cccccccccccccccc}
  \bigoplus_{i=1}^2 \theta_{1i}^{-1}& 0\\\
-V_H^{-1}W(\bigoplus_{i=1}^2 \theta_{1i}^{-1})  &V_{H}^{-1} \\
\end{array}
} \right],\quad
FV^{-1}= \left[ {%
\begin{array}{cccccccccccccccc}
 -UV_H^{-1}W(\bigoplus_{i=1}^2 \theta_{1i}^{-1}) & UV_H^{-1} \\
-F_HV_H^{-1}W(\bigoplus_{i=1}^2 \theta_{1i}^{-1})& F_HV_H^{-1}
\end{array}
} \right].
\end{equation*}
Since $\mathcal{M}^{-1}(FV^{-1})\mathcal{M}=\begin{bmatrix}
  0&UV_H^{-1}\\
0 & F_HV_H^{-1}-W(\bigoplus_{i=1}^2 \theta_{1i}^{-1})UV_H^{-1}
\end{bmatrix}$, where $\mathcal{M}=\begin{bmatrix}
  I_{2\times 2}&0\\
W(\bigoplus_{i=1}^2\theta_{1i}^{-1}) & I_{8\times 8}
\end{bmatrix}$, we have
\begin{equation}
R_0=\rho(FV^{-1})=\rho(F_HV_H^{-1}-W(\bigoplus_{i=1}^2\theta_{1i}^{-1})UV_H^{-1}).
\end{equation}
$R_0$ is a function of vertical and horizontal transmission parameters. Since  $F_HV_H^{-1}$ and  $-W(\bigoplus_{i=1}^2\theta_{1i}^{-1}) UV_H^{-1}$ are both nonnegative matrices, by Theorem \ref{theorem:spectral} in  appendix,
\begin{equation}\label{eq:roandroh}
  R_0 \geqslant \rho(F_HV_H^{-1}).
\end{equation}

\subsection{$R_0$ for multiple species in a general network}
\label{subsection:general model}
The model presented in Section \ref{subsubsec:two patches with vertical transmission}  is generalized to    model diseases transmitted among  all $h$ species  in node  $i \ (i=1, 2,
\cdots, n)$.
Suppose a disease is transmitted by species $k \ (k=1, 2, \cdots, h)$ vertically and horizontally  if $1 \leqslant  k \leqslant g$ and only horizontally otherwise.         The dynamical behavior is given by the  system with $4hn+2gn$ differential equations:
\allowdisplaybreaks
\begin{align}
 \frac{dP_{ki}}{dt} &=[r_{ki} -b_{ki} q_{ki}I_{ki}-\theta_{ki} P_{ki}]\delta (k) \label{equation:generalP}\\
 \frac{dQ_{ki}}{dt} &=[ b_{ki}q_{ki}I_{ki} -\theta_{ki} Q_{ki}]\delta (k)\\
 \frac{dS_{ki}}{dt}&=\theta_{ki} P_{ki}\delta (k)+r_{ki} (1-\delta (k))-\sum^h_{m=1}\beta_{mki}S_{ki}I_{mi}/N_{mi}
-d_{ki}S_{ki}+\sum^n_{j=1, j \neq i}\omega_{kji}S_{kj}-\sum^n_{j=1, j \neq i}\omega_{kij}S_{ki}\\
 \frac{dE_{ki}}{dt}&=\sum^h_{m=1}\beta_{mki}S_{ki}I_{mi}/N_{mi}
-\varepsilon_{ki}E_{ki}-d_{ki}E_{ki}+\sum^n_{j=1, j \neq i}\omega_{kji}E_{kj}-\sum^n_{j=1, j \neq i}\omega_{kij}E_{ki}\\
  \frac{\dif I_{ki}}{\dif t} &=\theta_{ki} Q_{ki}\delta (k)+\varepsilon_{ki}E_{ki}-\gamma_{ki}I_{ki}-d_{ki}I_{ki}+\sum^n_{j=1, j \neq i}\omega_{kji}I_{kj}-\sum^n_{j=1, j \neq i}\omega_{kij}I_{ki}\\
\frac{\dif R_{ki}}{\dif t} &=\gamma_{ki}I_{ki}-d_{ki}R_{ki}+\sum^n_{j=1, j \neq i}\omega_{kji}R_{kj}-\sum^n_{j=1, j \neq i}\omega_{kij}R_{ki}. \label{equation:generalR}
\end{align}
 The daily number of  species $k$ individuals infected by species  $m$   is $\beta_{mki}S_{ki}I_{mi}/N_{mi}$. The daily  numbers  of   species $k$ individuals  in compartment $J$ moving in and out of node $i$ are  $\sum^n_{j=1, j \neq i}\omega_{kji}J_{kj}$ and $\sum^n_{j=1, j \neq i}\omega_{kij}J_{ki}$, respectively. Other terms in the above equations have the same meanings as the corresponding ones in Section $\ref{subsubsection:no vertical transmission}$ (Equation (\ref{equation:susceptible}) to (\ref{equation:recovered})) and Section $\ref{subsubsec:two patches with vertical transmission}$ (Equation (\ref{equation:2.1.2P}) to (\ref{equation:2.1.2R})) except $\delta(k)$ defined below, which is used to differentiate  the horizontally-transmitting species and the species exhibiting both types of transmission.
\begin{equation*}
\delta (k)= \left\{ \begin{array}{rcl}
1 & \mbox{for}
&  1 \leq k \leq g,\\
0 & \mbox{for} & g+1 \leq k \leq h.
\end{array}\right.
\end{equation*}

 To compute $R_0$ using the next generation matrix method, we need to find matrices $\mathscr{F}$ and $\mathscr{V}$, omitted here due to large size.
In determining Jacobian matrices  $F$ and $V$, the infected variables are ordered by compartment, species, and  node index, i.e.,
\begin{align*}
 &Q_{11}, Q_{12}, \cdots, Q_{1n}, Q_{21}, Q_{22}, \cdots Q_{2n}, \cdots,  Q_{g1}, Q_{g2}, \cdots, Q_{gn},\\& E_{11}, E_{12}, \cdots,  E_{1n}, E_{21}, E_{22}, \cdots,  E_{2n}, \cdots, E_{h1}, E_{h2}, \cdots,  E_{hn},\\
& I_{11}, I_{12}, \cdots,  I_{1n}, I_{21}, I_{22}, \cdots,  I_{2n}, \cdots, I_{h1}, I_{h2}, \cdots,  I_{hn}.
\end{align*}
At DFE, $Q_{ki}=E_{ki}=I_{ki}=R_{ki}=0$, and $S_{ki}=N_{ki}$.
By Theorem \ref{theorem:N2} in  appendix,  a unique solution   $\left[ {%
\begin{array}{cccccccccccccccc}
N_{k1}^0 &  N_{k2}^0 \cdots N_{kn}^0
\end{array}
} \right] ^T$ exists.
Since incorporating multiple species in multiple nodes leads to  matrices $F$ and $V$ growing  very large, the computation of $R_0$ is simplified by decomposing the matrices into blocks, deriving block upper or lower triangular matrices as follows:
\begin{equation*}
F= \left[ {%
\begin{array}{cccccccccccccccc}
0 _{gn \times gn}& U_{gn \times2h n}\\
0_{2hn\times gn} &F_{H} \\
\end{array}
} \right],\ \ V= \left[ {%
\begin{array}{cccccccccccccccc}
\bigoplus_{k=1}^g( \bigoplus_{i=1}^n\theta_{ki})& 0_{gn\times 2hn}\\
W_{2hn\times gn} &V_{H} \\
\end{array}
} \right],
\end{equation*}
where
\begin{equation*}
F_H= \left[ {%
\begin{array}{cccccccccccccccc}
0_{hn \times hn}& \mathcal{A}_{hn \times hn}\\
0_{hn \times hn}&  0_{hn \times hn}\\
\end{array}
} \right],\quad
V_H
= \left[ {%
\begin{array}{cccccccccccccccc}
\bigoplus_{k=1}^h M_k& 0_{hn \times hn}\\
-\bigoplus_{k=1}^h (\bigoplus_{i=1}^n \varepsilon_{ki})&\bigoplus_{k=1}^h X_k
\end{array}
} \right],
\end{equation*}
\begin{equation*}
U
= \left[ {%
\begin{array}{cccccccccccccccc}
0_{gn\times hn} &\bigoplus_{k=1}^g(\bigoplus_{i=1}^n b_{ki}q_{ki}) &0_{gn\times (h-g)n }
\end{array}
} \right],\quad W
= \left[ {%
\begin{array}{cccccccccccccccc}
 0_{hn \times gn} \\
-\bigoplus_{k=1}^g(\bigoplus_{i=1}^n \theta_{ki}) \\
 0_{(h-g)n \times gn} \\
\end{array}
} \right].
\end{equation*}
The block matrix $\mathcal{A}$ in $F_H$ is written into an   $h \times h$ block matrix $\mathcal{A}=(\mathcal{A}_{km})$ and its $(k,m)$ entry is an $n\times n$ diagonal matrix
$\mathcal{A}_{km}=\bigoplus_{i=1}^n(\beta_{mki}\frac{S_{ki}^0}{N_{mi}^0}).$
 The matrices $M_k$ and $X_k$ are:
\begin{equation}
M_k= \left[ {%
\begin{array}{cccccccccccccccc}
\zeta_{k1}& -\omega_{k21} &  \cdots &   -\omega_{kn1}  \\
 -\omega_{k12} & \zeta_{k2}&  \cdots&  -\omega_{kn2} \\
\cdots & \cdots& \cdots &\cdots \\
 -\omega_{k1n}  & \cdots& \cdots &\zeta_{kn}\\
\end{array}
} \right],\quad {\rm and}\quad X_k= M_k+\bigoplus_{i=1}^n  (\gamma_{ki}- \varepsilon_{ki}), \label{matrix:generalMk}
\end{equation}
where $\zeta_{ki}=d_{ki}+\varepsilon_{ki}+\sum_{j=1, j \neq i}^n\omega_{kij}$.
Since  matrices $M_k$ and $X_k$ are invertible, according to Theorem \ref{theorem:X2M2}, $V_H$ and $V$ are invertible. It is easy to check:
\begin{equation}
V_H^{-1}=\left[ {%
\begin{array}{cccccccccccccccc}
\oplus _{k=1}^{h}M_k^{-1}&0\\
\oplus  _{k=1}^{h}\mathcal{Z}_k & \oplus  _{k=1}^{h} X_k^{-1}\\
\end{array}
} \right],\quad V^{-1}= \left[ {%
\begin{array}{cccccccccccccccc}
  \bigoplus_{k=1}^g (\bigoplus_{i=1}^n\theta_{ki}^{-1}) & 0_{gn\times 2hn}\\\
-V_H^{-1}W(\bigoplus_{k=1}^g(\bigoplus_{i=1}^n\theta_{ki}^{-1}) )&V_{H}^{-1} \\
\end{array}
} \right], \label{equation:VHinverse}
\end{equation}
where $\mathcal{Z}_k=X_k^{-1}(\oplus  _{i=1}^{n}\varepsilon_{ki}) M_k^{-1}$.
%
Similar to the derivation  in  Section \ref{sec2.2.4}, $R_0$ is:
\begin{equation}
 R_0=\rho(FV^{-1})=\rho(F_HV_H^{-1}-W(\bigoplus_{k=1}^g (\bigoplus_{i=1}^n\theta_{ki}^{-1}))UV_H^{-1}). \label{equation:R0general}
\end{equation}
Moreover, (\ref{eq:roandroh}) still holds.
If the lower bound $ \rho(F_HV_H^{-1})>1$, we can conclude that a network may be invaded without computing the upper bound or the exact value of $R_0$.

The term $F_HV_H^{-1}$ is related to horizontal transmission, and the term $ -W(\bigoplus_{k=1}^g (\bigoplus_{i=1}^n\theta_{ki}^{-1}))UV_H^{-1}$ is related to vertical transmission, making $R_0$ a function of vertical and horizontal transmission parameters.
  Generally, $R_0$ depends on demographic, disease and movement factors, proving too complicated to compute or analyze  \cite{Arino2007}. The complexity of computing $R_0$ using Equation $(\ref{equation:R0general})$ depends on a specific model for  a certain  disease. For the general model, we can only provide the formula of $R_0$ in Equation $(\ref{equation:R0general})$ and its lower bound in Inequality $(\ref{eq:roandroh})$.

In the following section, Equation $(\ref{equation:R0general})$  is applied  to  an RVF virus transmission meta-population model. Then, based on the assumptions for the RVF model,  we  compute $R_0$ using  Equation $(\ref{equation:R0general})$  and further derive  lower bound and upper bound, providing  insights  into the role of model parameters on $R_0$.
\section{The application of proposed method to  RVF meta-population model}
\label{section:RVFapplication}
Rift Valley fever  is an  emerging  mosquito-borne disease mainly affecting and colonizing domestic  ruminants and humans \cite{flick2005rift, Chevalier2010}.   Main vectors of RVF  include \it Aedes \rm  and \it Culex \rm  mosquitoes \cite{Chevalier2010}. Humans and ruminants are main hosts \cite{Chevalier2010}.  \it Aedes \rm mosquitoes are believed  to be initial source of RVF  outbreaks \cite{Crabtree2012}, since RVF virus-carrying eggs can survive in drought area soil for many years, later breeding infected mosquitoes in flooded habitats  \cite{Gerdes2002, Breiman2010}. Ruminants infected by mosquito bites  \cite{flick2005rift} can transmit RVF virus to Aedes feeding on them as blood meals  \cite{Chevalier2010}.
Culex mosquitoes also amplify RVF virus transmission by ingesting blood from infected ruminants  \cite{flick2005rift}. Most humans acquire RVF virus infection when bitten by infected mosquitoes or during contact with  body fluid  of infected ruminants  \cite{Linthicum1999}.
Next, we derive $R_0$ for an  RVF meta-population  model to study the role of parameters and networks on the reproduction number.
\subsection{The network-based  RVF  meta-population model}
\label{subsection:model}
In this section, the general model in Equations $(\ref{equation:generalP})$ to $(\ref{equation:generalR} )$ of  Section $\ref{subsection:general model}$  is applied to study  the dynamics of   RVF virus transmission with $h=4$, $g=1$.  \it Aedes \rm  and \it Culex \rm mosquito vectors are considered in the model,   as are livestock and human hosts.  The  RVF model is less complex than the general model presented in Equations $(\ref{equation:generalP})$ to $(\ref{equation:generalR})$.  Here, we assume only livestock
 can move in and out of nodes, and all mosquitoes do not recover. We consider
 disease-induced mortality for livestock and humans, and  carrying capacity for mosquitoes and humans. Due to lack of transmission by humans or direct intra-species transmission, this  RVF model contains fewer infection terms than those in the general model. See  appendix for the full model (Equations $(\ref{equation:P1})$ to $(\ref{equation:N4})$) and relative parameters (Table $\ref{table:parameters})$. The number of species $k$ individuals  $ (k=1, 2, 3,  4)$  from node $i \ (i=1, 2, \cdots,  n)$ in compartment $J$ is represented by $J_{ki}$, where $k=1$  (resp. 2, 3, 4) represents  \it  Aedes \rm mosquitoes (resp.  livestock,   \it Culex \rm mosquitoes, and  humans).   The parameter $r_{2i}$ is the number of  livestock born daily   in node $i \  (i=1, 2, \cdots,n)$. The daily numbers  of new born \it Aedes \rm mosquitoes, \it Culex  \rm mosquitoes, and humans are $b_{ki}N_{ki}$. A node index is added at the end of the subscript of a parameter only when referring to  a parameter for a specific node.  For example, $\beta_{12i}$ represents the contact rate from \it Aedes \rm mosquitoes ($k=1$) to livestock ($k=2$) in node $i$.

\subsection{The computation  of  $R_0$  for   RVF }

The explicit expression of $R_0$ in Equation $(\ref{equation:R0general})$ is applied to the RVF meta-population model. The above assumptions allow us to obtain the lower and upper bounds of $R_0$.
\subsubsection{Explicit  expression of  $R_0$ for RVF}
First, we check if  a unique solution  $N_{ki}^0$   exists.
At  DFE, $E_{ki}^0=I_{ki}^0=R_{ki}^0=0$.
By computation,
$S^0_{ki} =N^0_{ki}  =\frac{b_{ki}K_k}{d_{ki}}$ for $k=1, 3, 4$, where $K_k$ is the carrying capacity of species $k$. This  is a special case of the model  for Theorem \ref{theorem:N2}, which generates a unique nonnegative solution for the total number of livestock in node $i$ at DFE denoted by:
$\left[ {%
\begin{array}{cccccccccccccccc}
N_{21}^0 & N_{22}^0 &  \cdots  & N_{2n}^0
\end{array}
} \right]^T
$.

By (\ref{Jacobian}), the Jacobian matrices for the  RVF model are:
\begin{equation*}
F= \left[ {%
\begin{array}{cccccccccccccccc}
0 _{n \times n}& U_{n \times 8n}\\
0_{8n\times n} &F_{H} \\
\end{array}
} \right],\quad
V= \left[ {%
\begin{array}{cccccccccccccccc}
\oplus  _{i=1}^{n}\theta_{1i} & 0_{n \times 8n}\\
W_{8n\times n} &V_{H} \\
\end{array}
} \right].
\end{equation*}
Each component of the $R_0$ formula is computed as follows:
\begin{equation}\label{equation:FH}
F_H= \left[ {%
\begin{array}{cccccccccccccccc}
0_{4n \times 4n}& \mathcal{A}_{4n \times 4n}\\
0_{4n \times 4n}&  0_{4n \times 4n}\\
\end{array}
} \right],\quad
V_H= \left[ {%
\begin{array}{cccccccccccccccc}
\bigoplus_{k=1}^4 M_k& 0_{4n \times 4n}\\
-(\oplus  _{k=1}^{4} (\oplus  _{i=1}^{n}\varepsilon_{ki})) _{4n \times 4n}&\bigoplus_{k=1}^4 X_k
\end{array}
} \right].
\end{equation}
\begin{equation}\label{equation:U}
U
= \left[ {%
\begin{array}{cccccccccccccccc}
0_{n\times 4n}&\oplus  _{i=1}^{n}(b_{1i}q_{1i})&0_{n\times 3n} \\
\end{array}
} \right],\quad
W
= \left[ {%
\begin{array}{cccccccccccccccc}
0_{4n\times n} \\
-(\oplus  _{i=1}^{n} \theta_{1i})\\
0_{3n\times n}\\
\end{array}
} \right].
\end{equation}
\begin{equation}
\mathcal{A}
= \left[ {%
\begin{array}{cccccccccccccccc}
 0& \mathcal{A}_{12}  & 0 & 0\\
\mathcal{A}_{21}& 0 &\mathcal{A}_{23}& 0\\
 0 & \mathcal{A}_{32}  & 0 & 0\\
 \mathcal{A}_{41} & \mathcal{A}_{42}  & \mathcal{A}_{43}& 0\\
\end{array}
} \right], \label{equation:mathcalA}
\end{equation}
\begin{align*}
\mathcal{A}_{12}&= \oplus _{i=1}^{n}\beta_{21i}\frac{S_{1i}^0}{N_{2i}^0},\ \
\mathcal{A}_{21}=\oplus _{i=1}^{n}\beta_{12i}\frac{S_{2i}^0}{N_{1i}^0},\ \
\mathcal{A}_{23}=\oplus _{i=1}^{n}\beta_{32i} \frac{S_{2i}^0}{N_{3i}^0},\ \
\mathcal{A}_{32}=\oplus _{i=1}^{n}\beta_{23i}\frac{S_{3i}^0}{N_{2i}^0},\\
\mathcal{A}_{41}&=\oplus _{i=1}^{n}\beta_{14i}\frac{S_{4i}^0}{N_{1i}^0},\ \
\mathcal{A}_{42}= \oplus _{i=1}^{n}\beta_{24i}\frac{S_{4i}^0}{N_{2i}^0},\ \
\mathcal{A}_{43}=\oplus _{i=1}^{n}\beta_{34i}\frac{S_{4i}^0}{N_{3i}^0}.
\end{align*}

The matrices $V_H^{-1}$ and $V^{-1}$ are in Equation $(\ref{equation:VHinverse})$ 
 with g=1 and h=4, respectively.
Below,  matrices $M_k$ and $X_k$ relate to \it Aedes \rm mosquitoes,  livestock, \it Culex \rm mosquitoes,  and humans with $k=1,2,3,4$, respectively.
$$\begin{array}{lll}
M_1= \oplus _{i=1}^{n}(\frac{d_{1i}N_{1i}^0}{K_1}+\varepsilon_{1i}), &  X_1=M_1- \oplus _{i=1}^{n}\varepsilon_{1i}, \\
M_3= \oplus _{i=1}^{n}(\frac{d_{3i}N_{3i}^0}{K_3}+\varepsilon_{3i}),&  X_3=M_3- \oplus _{i=1}^{n}  \varepsilon_{3i},\\
M_4= \oplus _{i=1}^{n}(\frac{d_{4i}N_{4i}^0}{K_4}+\varepsilon_{4i}), & X_4=M_4- \oplus _{i=1}^{n}  \varepsilon_{4i},\\
\end{array}$$
\begin{equation*}
M_2= \left[ {%
\begin{array}{cccccccccccccccc}
\zeta_{21}& -\omega_{221} &  \cdots &   -\omega_{2n1}  \\
 -\omega_{212}& \zeta_{22}&  \cdots&  -\omega_{2n2}  \\
\cdots & \cdots& \cdots &\cdots \\
 -\omega_{21n} & -\omega_{22n} & \cdots &\zeta_{2n}\\
\end{array}
} \right], \quad
X_2= M_2+ \oplus _{i=1}^{n}  (\gamma_{2i}+\mu_{2i}-\varepsilon_{2i}).
\end{equation*}
 The reproduction number, $R_0$ can be computed by plugging the above terms into Equation (\ref{equation:R0general}).   Typically $R_0$ for a meta-population model is complicated \cite{Arino2009}. Deriving some bounds on the value of $R_0$ can be helpful \cite{Arino2009}.  In the following section, we derive lower and upper bounds for $R_0$.
\subsubsection{Deriving lower bound and upper bound for $R_0$ }
Bounds of  $R_0$ are derived in many articles, among which  are some  following examples. Gao and  Ruan  present bounds of $R_0$ for an SIS patch model  \cite{Gao2011} investigating effects of media coverage and human movement on the spread of infectious diseases, as well as a malaria model \cite{Gao2012paper}.    Hsieh, Driessche, and Wang \cite{Hsieh2007} derive bounds of $R_0$,   describing
the relationship between the  reproduction numbers for the isolated $i^{th}$ patch  and
for the system.   Salmani and Driessche  \cite{Salmani2006}  derive bounds for an SEIRS patch model. Arino \cite{Arino2009} presents bounds of $R_0$  for patch models  considering multiple species.  The reproduction number for an averaging process of mixed individuals or groups is estimated to be smaller than or equal to the  reproduction number before mixing \cite{Adler1992}.
We derive  bounds of $R_0$ for RVF  meta-population model in this section. In the following, we shall state main results and prove them in  appendix.
\begin{thm} \label{theorem:R0} Consider the model presented in Section \ref{subsection:model} (Equations $(\ref{equation:P1})$ to $(\ref{equation:N4})$), we obtain
\begin{equation}
\rho(F_HV_H^{-1}) \leqslant  R_0 \leqslant   \rho(F_HV_H^{-1})+\max_i(q_{1i}). \label{equation:R0bounds}
\end{equation}
\end{thm}
The difference between the lower and upper bounds is $\max_i(q_{1i})$ with lower bound $\rho(F_HV_H^{-1})$ computed by Equation $(\ref{equation:rhoRHVHinverse})$.

\begin{thm} \label{theorem:FHVHinverse}
For the model in Section \ref{subsection:model} (Equations $(\ref{equation:P1})$ to $(\ref{equation:N4})$), assume  $\varepsilon_{2i}=\varepsilon_{2}$ for all $i$, then
\begin{equation}
  \sqrt{\min_i(\chi_i)\rho(X_2^{-1}M_2^{-1})} \leqslant R_0 \leqslant \sqrt{\max_i(\chi_i) \rho(X_2^{-1}M_2^{-1})}+\max_i(q_{1i}),  \label{equation:rhoFHVHinverse1}
\end{equation}
where
 \begin{equation}\label{chii}
\chi_i =\frac{\varepsilon_{1i}\varepsilon_{2}\beta_{12i}\beta_{21i}}{b_{1i}(b_{1i}+\varepsilon_{1i})}+
  \frac{\varepsilon_{2}\varepsilon_{3i}\beta_{32i}\beta_{23i}}{b_{3i}(b_{3i}+\varepsilon_{3i})}.
 \end{equation}
\end{thm}

The difference between the lower bound and the upper bound in a network with heterogeneous corresponding parameters across nodes is larger than that in Inequality  $(\ref{equation:R0bounds})$.

\begin{cor}\label{cor1}
Suppose for  all $i$, birth  and incubation rates in mosquitoes and  livestock, contact rates between  livestock and  mosquitoes   are homogeneous for different nodes, i.e.,
\begin{equation}
b_{1i}=b_1, \  b_{3i}=b_3,\   \varepsilon_{1i}=\varepsilon_1,\  \varepsilon_{2i}=\varepsilon_2,  \  \varepsilon_{3i}=\varepsilon_3, \
 \beta_{12i}=\beta_{12},\   \beta_{21i}=\beta_{21},\   \beta_{23i}=\beta_{23},\   \beta_{32i}=\beta_{32}. \label{parameter2}
\end{equation}
Then
\begin{equation}
\sqrt{\chi \rho(X_2^{-1}M_2^{-1})} \leqslant  R_0 \leqslant \sqrt{\chi \rho(X_2^{-1}M_2^{-1})}+\max_i(q_{1i}). \label{equation:boundsofFVinverse}
\end{equation}
where
\begin{equation}
\chi=\frac{\varepsilon_1\varepsilon_2\beta_{12}\beta_{21}}{b_1(b_1+\varepsilon_1)}+\frac{\varepsilon_2\varepsilon_3\beta_{32}\beta_{23}}{b_3(b_3+\varepsilon_3)}. \label{equation:chi}
\end{equation}
\end{cor}

 \begin{thm}\label{theorem:heterogeneousparameter}
Under the condition  of Theorem \ref{theorem:FHVHinverse}, $R_0$ can be estimated by the following   inequality:
 \begin{align}
\sqrt{ \frac{\min_i{(\chi_i)}}{\max_i{(d_{2i}+\varepsilon_{2})\max_i(d_{2i}+\gamma_{2i}+\mu_{2i})}}}  \leqslant  R_0 \leqslant \sqrt{ \frac{\max_i{(\chi_i)}}{\min_i{(d_{2i}+\varepsilon_{2})\min_i(d_{2i}+\gamma_{2i}+\mu_{2i})}}}+\max_i(q_{1i}). \label{equation:boundsestimation}
\end{align}
 \end{thm}

If the differences between $\min_i{(\chi_i)} $ and $\max_i{(\chi_i)}$, $\min_i{(d_{2i}+\varepsilon_{2}})$   and  $\max_i{(d_{2i}+\varepsilon_{2}})$,   $\min_i(d_{2i}+\gamma_{2i}+\mu_{2i})$ and $\max_i(d_{2i}+\gamma_{2i}+\mu_{2i})$  are large, then the difference between the lower bound and the upper bound may be large.   However, the scalar lower bound and upper bound are easily computed.  Moreover, if the lower bound is greater than $1$, we can conclude that the network may be invaded without computing  $R_0$ or its upper bound.

 \begin{cor}\label{cor:homogeneous} Based on the condition of Corollary $\ref{cor1}$, we further assume that for all $ i$, the death rate,  mortality rate, and recovery rate in  livestock, and transovarial transmission rate in \it Aedes  mosquitoes are homogeneous for all nodes, i.e.,
\begin{equation}
d_{2i}=d_2, \  \ \mu_{2i}=\mu_2 , \  \  \gamma_{2i}=\gamma_{2}, \  \  q_{1i}=q_{1}.  \label{parameter3}
\end{equation}
Then  \rm \begin{equation}
\sqrt{ \frac{\chi  }{(d_2+\varepsilon_2)(d_2+\gamma_2+\mu_2)}}\leqslant  R_0  \leqslant \sqrt{ \frac{\chi  }{(d_2+\varepsilon_2)(d_2+\gamma_2+\mu_2)}}+q_{1} . \label{inequality:homogeneousbounds}
\end{equation}
 \end{cor}

In this case,  the lower and upper bounds  of $R_0$ correspond to the bounds for homogeneous populations presented in  \cite{LingXue2010} and  are tight \cite{LingXue2010}. Clearly, $R_0$ for horizontal transmission,
\begin{equation}R_0^H=\sqrt{
\frac{\varepsilon_2}{(b_2+\varepsilon_2)
(b_2+\gamma_2+\mu_2)}
\Big[\frac{\varepsilon_1\beta_{12}\beta_{21}}{b_1
 (b_1+\varepsilon_1)}
+\frac{\varepsilon_3\beta_{32}\beta_{23}}{b_3(b_3+\varepsilon_3)}
\Big] }\label{equation:R0HRVF}, \end{equation}
 does not depend on  livestock movement rates.  Only bounds for $R_0$ can theoretically be obtained. Based on numerical simulation results, we conjecture that, given the conditions for  Corollary \ref{cor:homogeneous},  $R_0$ does not depend on livestock movement rates.

\subsubsection{Tightness of  bounds for $R_0$}

A one hundred-node network with heterogeneous corresponding parameters among nodes is built to study the tightness of bounds.
 We uniformly distribute disease parameters for each node during  one hundred runs within their respective ranges, given  in Table \ref{table:parameters}. Then,  $R_0$ is numerically computed according to   Equation $(\ref{equation:R0general})$.  Lower and upper bounds of $R_0$ are computed according to   Inequality  $(\ref{equation:R0bounds})$  in Theorem  $\ref{theorem:R0}$. The reproduction number for horizontal transmission  is computed according to Equation $(\ref{equation:rhoRHVHinverse})$. The lower bound of $R_0$ (denoted by $R_0^L$)    versus   $R_0$  in each run  is shown in Figure \ref{fig:lowerbound}, and the upper bound  of $R_0$  (denoted by $R_0^U$) versus $R_0$ in each run is shown in Figure \ref{fig:upperbound}. In each run, the upper bound is slightly greater and the lower bound is slightly smaller than $R_0$.  With the same network and the same set of parameters, the lower and upper bounds of $R_0$  are computed using Inequality    $(\ref{equation:rhoFHVHinverse1})$.   The lower bound versus exact $R_0$  is shown in Figure $\ref{fig:lowerboundhe}$, and the upper bound versus exact $R_0$  is   shown in Figure \ref{fig:upperboundhe}. The bounds obtained by  Inequality $(\ref{equation:rhoFHVHinverse1})$  in Theorem  $\ref{theorem:FHVHinverse}$  are less tight than those obtained by Inequality  $(\ref{equation:R0bounds})$   in Theorem  $\ref{theorem:R0}$, as $\rho(F_HV_H^{-1})$ is estimated by computing the spectral radius of a smaller size matrix. The bounds obtained by Inequality  $(\ref{equation:boundsestimation})$ in Theorem  $\ref{theorem:heterogeneousparameter}$ can be even looser because $\rho(X_2^{-1}M_2^{-1})$ is simply  estimated by scalars.

The above  bounds are for heterogeneous networks. The bounds in Corollary $\ref{cor:homogeneous}$ (see Inequality $(\ref{inequality:homogeneousbounds})$) apply to homogeneous networks, where the difference between the lower bound and the upper bound is  the largest transovarial transmission rate of  \it Aedes \rm mosquitoes across nodes.
\begin{small}
\begin{figure}[!h]
\centering
\subfigure[The reproduction number and its lower  bound with  heterogeneous parameters.]{
\label{fig:lowerbound}
\includegraphics[angle=0,width=6cm,height=6cm]{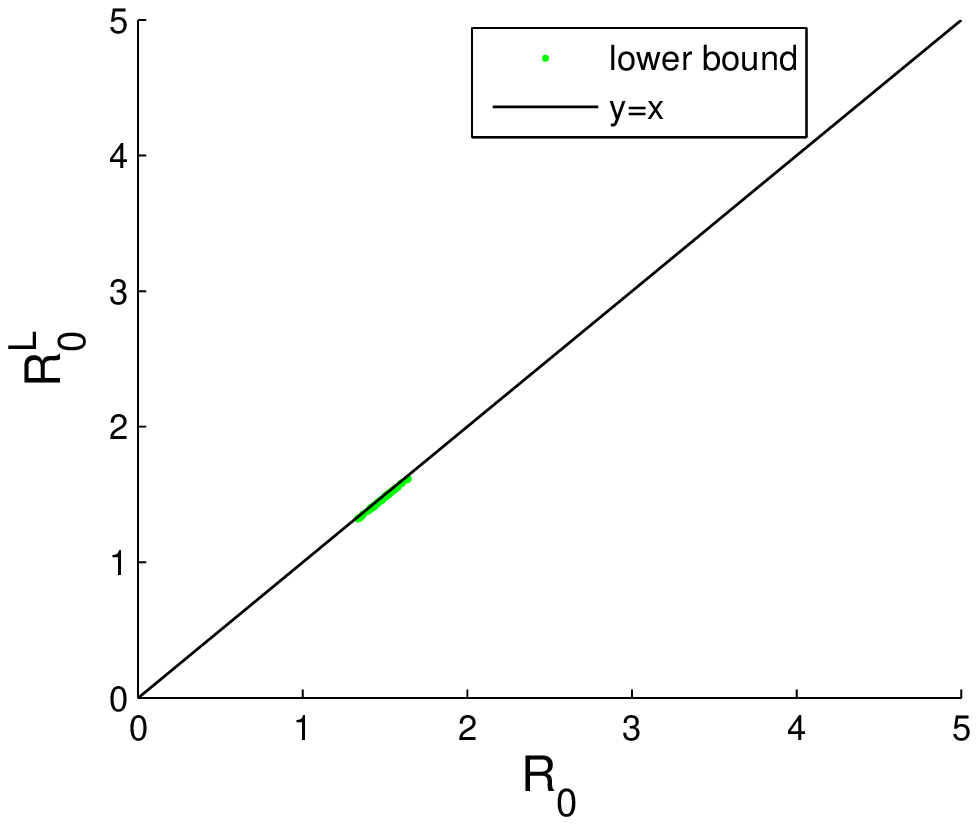}}
\hspace{0.1in}
\subfigure[The reproduction number and its upper  bound with  heterogeneous  parameters.]{
\label{fig:upperbound}
\includegraphics[angle=0,width=6cm,height=6cm]{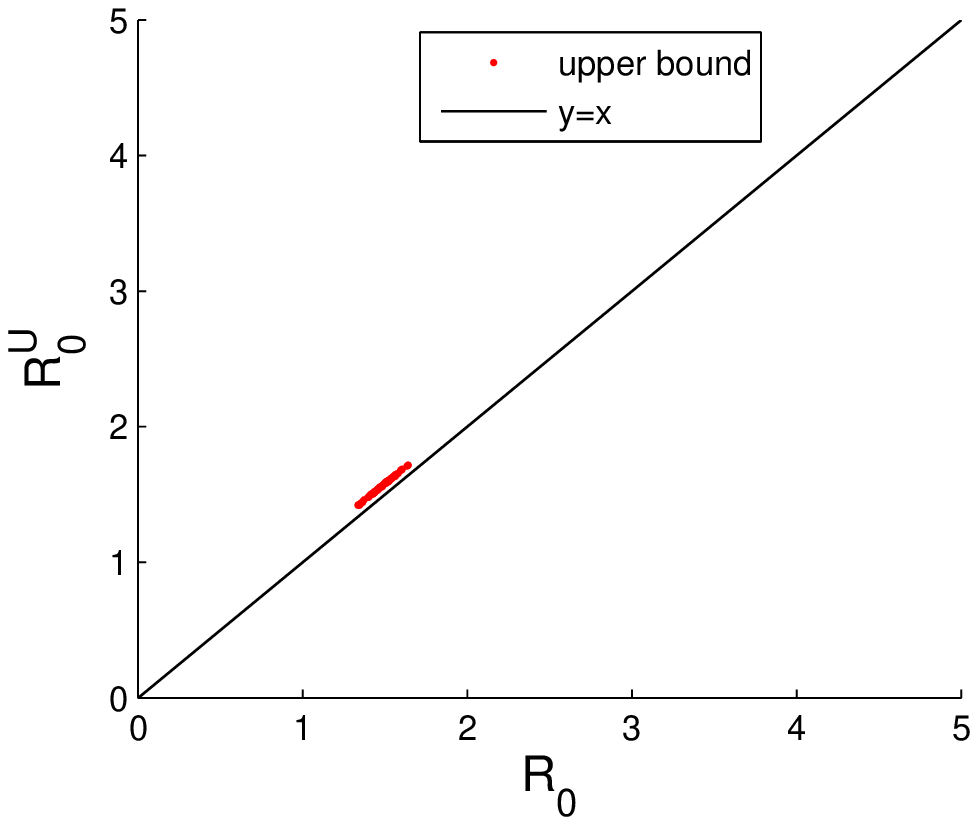}}
\hspace{0.1in}
\caption{The reproduction number and its  lower and upper  bounds  computed using Theorem $\ref{theorem:R0}$ for one hundred simulation runs in  one hundred-node heterogeneous networks. }
\label{fig:ApproximateR0}
\end{figure}
\end{small}

\begin{small}
\begin{figure}[!h]
\centering
\subfigure[The reproduction number and its lower  bound with heterogeneous parameters.]{
\label{fig:lowerboundhe}
\includegraphics[angle=0,width=6cm,height=6cm]{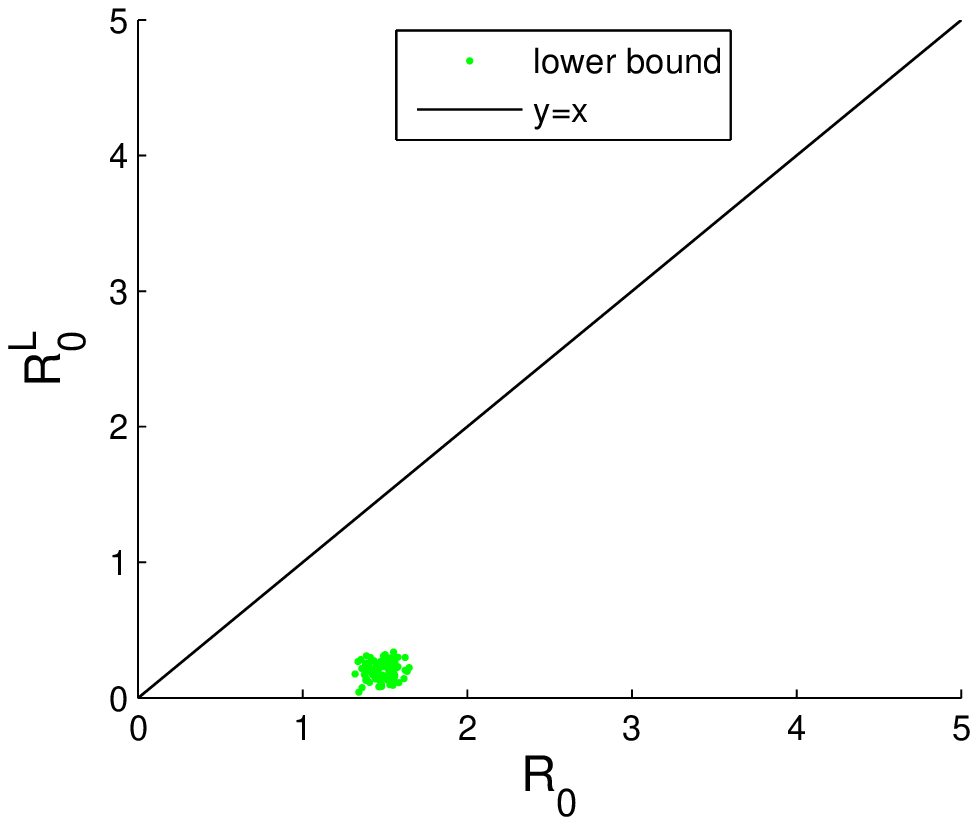}}
\hspace{0.1in}
\subfigure[The reproduction number and its upper  bound with heterogeneous parameters.]{
\label{fig:upperboundhe}
\includegraphics[angle=0,width=6cm,height=6cm]{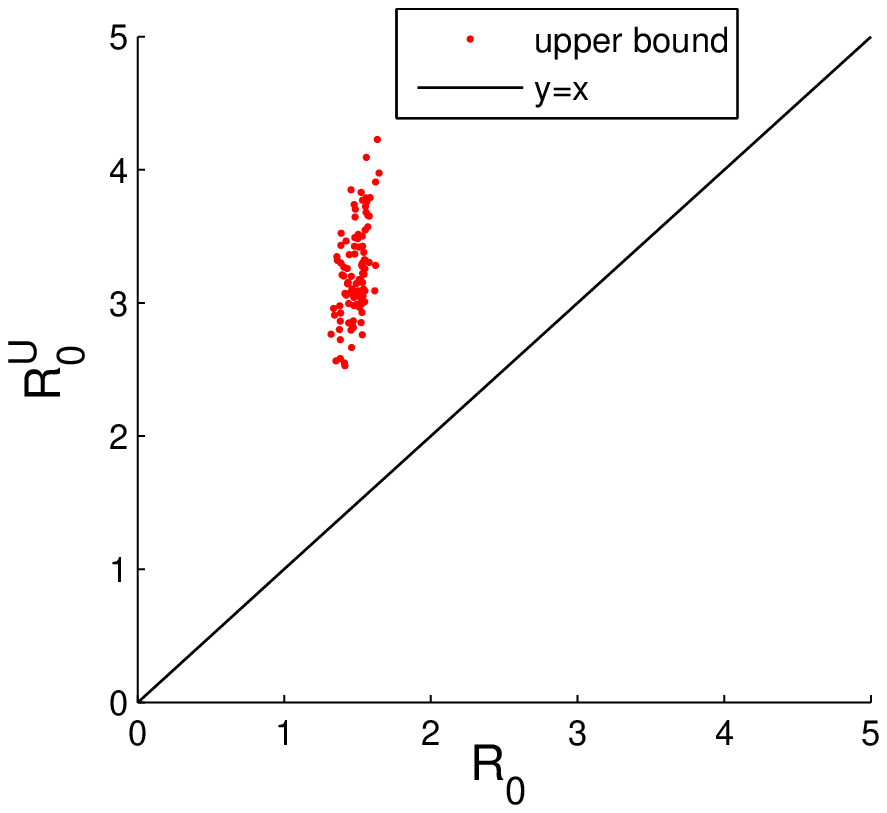}}
\hspace{0.1in}
\caption{The reproduction number and its  lower and upper  bounds  computed using Theorem $\ref{theorem:FHVHinverse}$ for one hundred simulation runs in  one hundred-node heterogeneous networks.}
\label{fig:ApproximateR0he}
\end{figure}
\end{small}

\subsection{Assessing the role of parameters on $R_0$}
As an example, a two-node network demonstrates how bounds of $R_0$ alter with livestock movement rates, if parameters  $d_{2i}$,  $\gamma_{2i}$, and $\mu_{2i}$ are heterogeneous, i.e., at least one of inequalities $d_{2i} \neq d_{2j} $,  $\gamma_{2i}  \neq \gamma_{2j}$, $\mu_{2i}  \neq \mu_{2j}$ holds for different $i$ and $j$. In this example, $M_2$ corresponds to  the one in Equation (\ref{M1M2}) and $X_2=M_2+\oplus_{i=1}^2(\gamma_{2i}+\mu_{2i}-\varepsilon_{2i})$.  Since $X_2$, $M_2$ are both diagonal dominant matrices, by Theorem $\ref{theorem:X2M2}$, $M_2^{-1}$ and $X_2^{-1}$ are both nonnegative matrices.

According to  Proposition $4.3$ in  \cite{Gao2012paper}, $\rho(X_2^{-1}M_2^{-1})$ is decreasing in $\omega_{212}$ if
$$\omega_{212}(a_2-a_1)>(a_1c_1-a_2c_2)-(a_2-a_1)\omega_{221}$$
 and increasing otherwise, where $a_1=d_{21}+\varepsilon_{21},  a_2=d_{22}+\varepsilon_{22}$, $c_1=d_{21}+\gamma_{21}+\mu_{21}$ and $c_2=d_{22}+\gamma_{22}+\mu_{22}$. In the case that $a_1=a_2$, $\rho(X_2^{-1}M_2^{-1})$ is decreasing in $\omega_{212}$ if $c_2>c_1$ and increasing otherwise. If $a_1 \neq a_2$, $\omega_{212}^*:=\frac{a_1c_1-a_2c_2}{a_2-a_1}-\omega_{221}$ is a critical point of $\rho(X_2^{-1}M_2^{-1})$.
 Moreover, $\rho(X_2^{-1}M_2^{-1})$ reaches the maximum value at $\omega_{212}^*$ if $a_2>a_1$ and the minimum value at $\omega_{212}^*$ otherwise.

To evaluate the impact of networks with corresponding homogeneous parameters across all nodes on  the value of $R_0$ computed using Equation $(\ref{equation:R0general})$, we construct three networks with three, four, and one hundred nodes, respectively. Simulation runs with varying livestock movement rates, and  parameters in  $(\ref{parameter2})$  and $(\ref{parameter3})$   held constant  and homogeneous across nodes  showed  $R_0$ is not affected by livestock movement rates during one hundred runs per network.    Moreover, the values and bounds of $R_0$ obtained through  numerical simulations are the same for networks with three, four, and one hundred nodes.  Through extensive numerical simulations,  we have observed that  $R_0$ does not depend on livestock movement rates or the number of nodes in a network when    $(\ref{parameter2})$ and $(\ref{parameter3})$  hold.

 We run scenarios (see Table $\ref{table:scenarios}$)  one hundred times for each four-node  network  to study the impact of livestock movement rates on $R_0$.  During one hundred realizations for each scenario, we increase livestock movement rates while keeping remaining parameters constant and homogeneous across all nodes.    In Scenario $1$, we set  contact rates $\beta_{12}$,   $\beta_{21}$,
 $\beta_{23}$, and  $\beta_{32}$ for node $i$  larger than respective parameters for  node $j$ ($i>j, \ i,  j=1, 2, 3, 4$).
During each run, $R_0$ increases while increasing livestock movement rates from node  $j$ to node $i$,  $\omega_{2ji}$,  and decreases while increasing livestock movement rates from node  $i$ to node $j$,  $\omega_{2ij}$ (see  Figure \ref{fig:movestol} and \ref{fig:moveltos}, respectively).
In Scenario $2$, under setting $d_{2i}>d_{2j}$, $R_0$ decreases when  $\omega_{2ji}$  increases, and  increases
when  $\omega_{2ij}$ increases (see Figure  \ref{fig:movedeathltos} and \ref{fig:movedeathstol},
respectively). With livestock recovery rates $\gamma_{2i}>\gamma_{2j}$ in Scenario $3$,  $R_0$ decreases when
 $\omega_{2ji}$ increases, and  increases when   $\omega_{2ij}$ increases (see Figure  \ref{fig:moverecoverltos}
and  \ref{fig:moverecoverlsol}, respectively). Similarly, when livestock mortality rates  $\mu_{2i}>\mu_{2j}$ in Scenario $4$, $R_0$
 decreases when $\omega_{2ji}$ increases, and increases with larger  $\omega_{2ij}$ (see
  Figure \ref{fig:movemortalityltos} and  \ref{fig:movesmortalitysol}, respectively).
Tuning the parameters in above scenarios yields $R_0$  from below $1$ to above $1$. As a consequence, livestock movement rates are important in either leading to an epidemic outbreak or epidemic burnout.

\begin{table}
\centering
\begin{tabular}{|p{15pt}|p{240pt}|p{100pt}|p{45pt}|}
 \hline
No. &parameter& livestock movement rates &$R_0$\\
\hline
$1$  & $\beta_{12i}>\beta_{12j}$, \ $\beta_{21i}>\beta_{21j}$, \ $\beta_{23i}>\beta_{23j}$, \  $\beta_{32i}>\beta_{32j} $  &  $\omega_{2ji}$ increases& increases\\
 &    &$\omega_{2ij}$ increases& decreases\\
$2$  & $d_{2i}>d_{2j} $ & $\omega_{ji}^2 $ increases&decreases\\
 & & $\omega_{2ij}$ increases&increases\\
$3$& $\gamma_{2i}>\gamma_{2j}  $&$\omega_{2ji}$ increases &decreases\\
&&$\omega_{2ij}$ increases &increases\\
$4$ &$\mu_{2i}>\mu_{2j} $ &$\omega_{2ji}$ increases &decreases \\
 & &$\omega_{2ij}$ increases &increases \\
\hline
\end{tabular}
\caption{Different scenarios for numerical simulations in four-node networks. Other parameters are kept the same and homogeneous across all nodes  during all realizations. The superscripts $i, j= 1, 2, 3, 4$ and $i>j$.}
\label{table:scenarios}
\end{table}

\begin{small}
\begin{figure}[!h]
\centering
\subfigure[As the  livestock movement rate from node $j$ to node $i$ ($\omega_{2ji}$) increases when $\beta_{12i}>\beta_{12j}$, $\beta_{21i}>\beta_{21j}$, $\beta_{23i}>\beta_{23j}$, and  $\beta_{23i}>\beta_{23j}$,  $R_0$ increases.]{
\label{fig:movestol}
\includegraphics[angle=0,width=6cm,height=5cm]{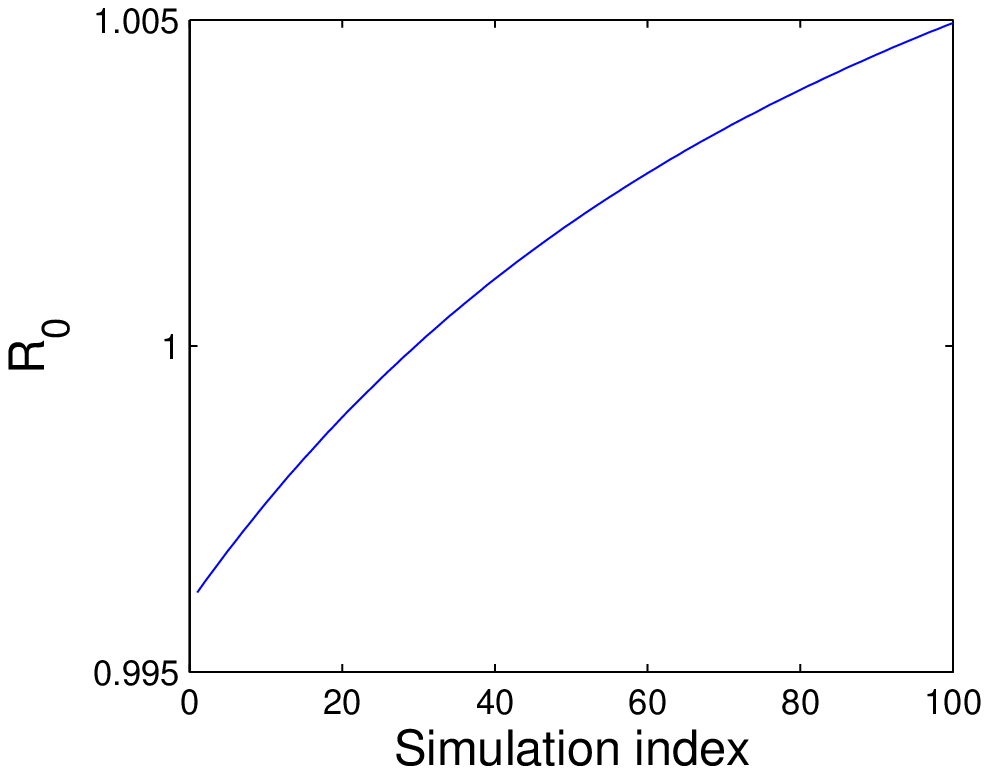}}
\hspace{0.1in}
\subfigure[As the  livestock movement rate from node $i$ to node $j$ ($\omega_{2ij}$) increases when $\beta_{12i}>\beta_{12j}$, $\beta_{21i}>\beta_{21j}$, $\beta_{23i}>\beta_{23j}$, and  $\beta_{23i}>\beta_{23j}$, $R_0$ decreases.]{
\label{fig:moveltos}
\includegraphics[angle=0,width=6cm,height=5cm]{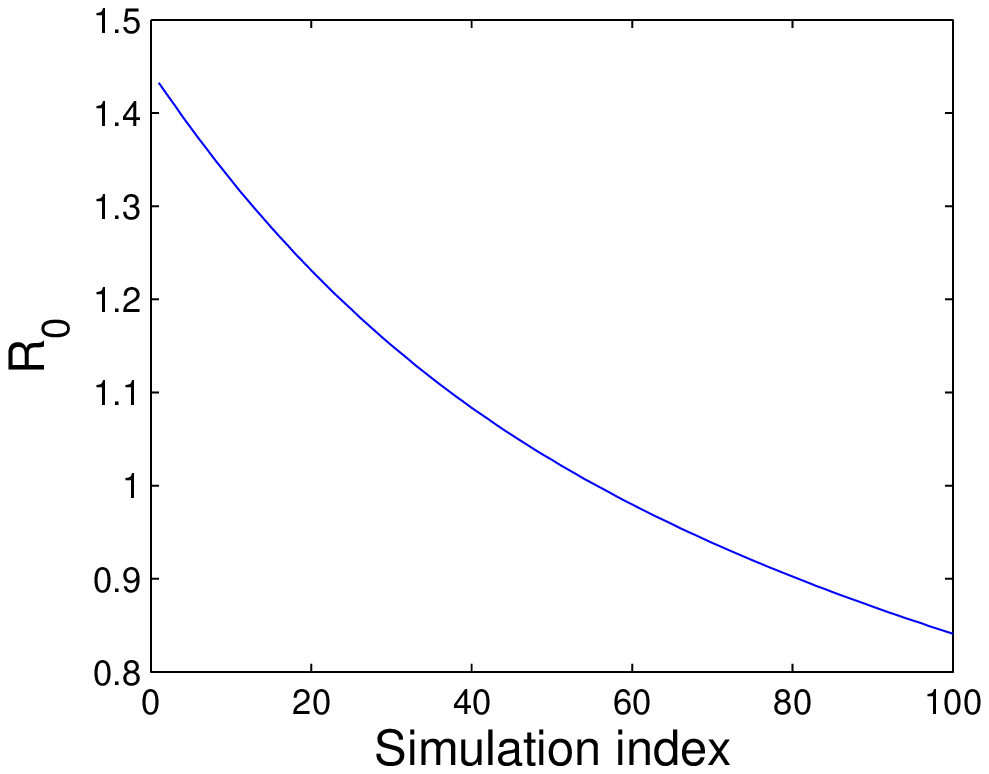}}
\hspace{0.1in}
\caption{The reproduction number for  four-node networks   with different contact rates during one hundred runs. }
\label{fig:differentcontactrates}
\end{figure}

\begin{figure}[!h]
\centering
\subfigure[As the  livestock movement rate  from node $j$ to node $i$ ($\omega_{2ji}$) increases when $d_{2i}>d_{2j} $, $R_0$ decreases.]{
\label{fig:movedeathltos}
\includegraphics[angle=0,width=6cm,height=5cm]{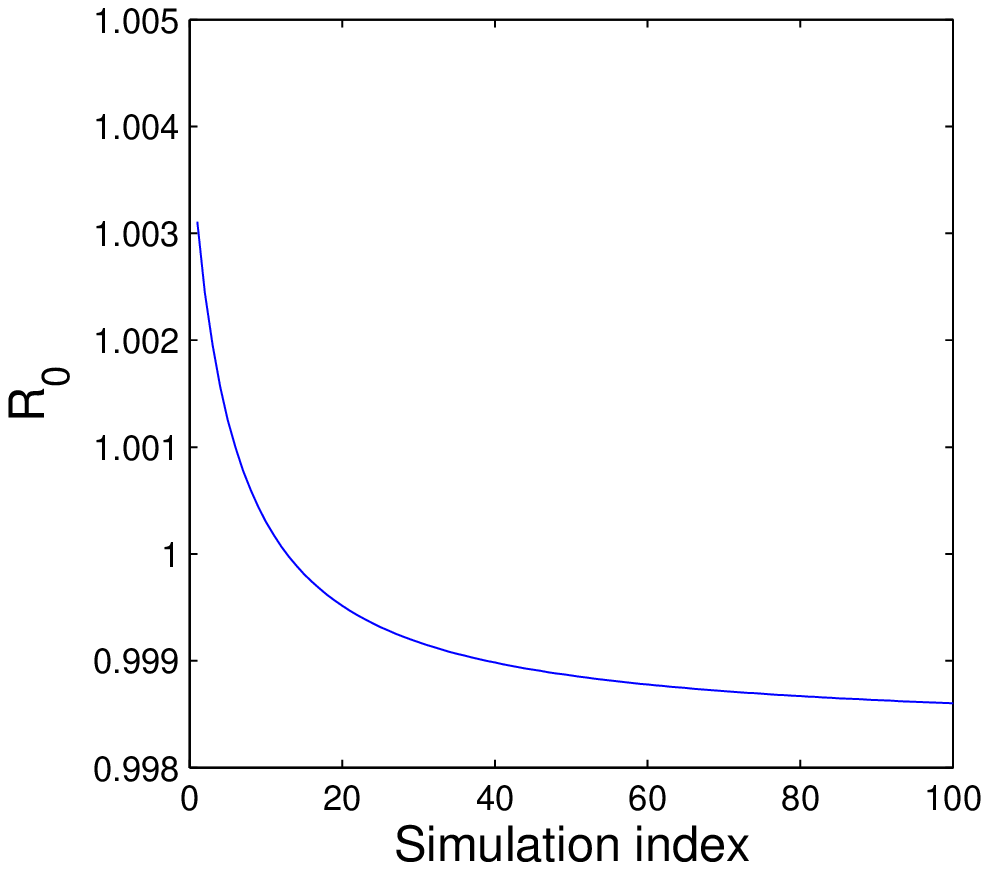}}
\hspace{0.1in}
\subfigure[As the  livestock movement rate from node $i$ to node $j$ ($\omega_{2ij}$) increases when $d_{2i}>d_{2j} $, $R_0$ increases.]{
\label{fig:movedeathstol}
\includegraphics[angle=0,width=6cm,height=5cm]{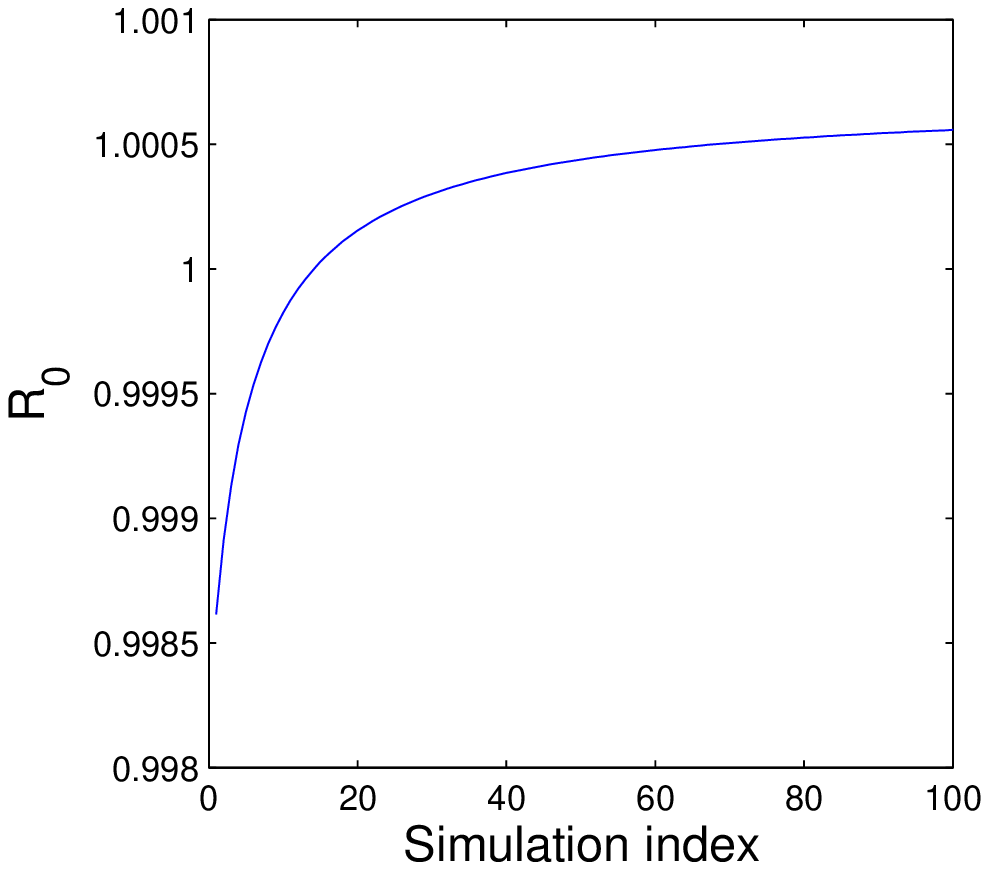}}
\hspace{0.1in}
\caption{The reproduction number for  four-node networks  with different livestock death  rates during one hundred runs.}
\label{fig:differentdeathrates}
\end{figure}
\begin{figure}[!h]
\centering
\subfigure[As the livestock movement rate from node $j$ to node $i$ ($\omega_{2ji}$) increases when $\gamma_{2i}>\gamma_{2j} $,  $R_0$ decreases.]{
\label{fig:moverecoverltos}
\includegraphics[angle=0,width=6cm,height=5cm]{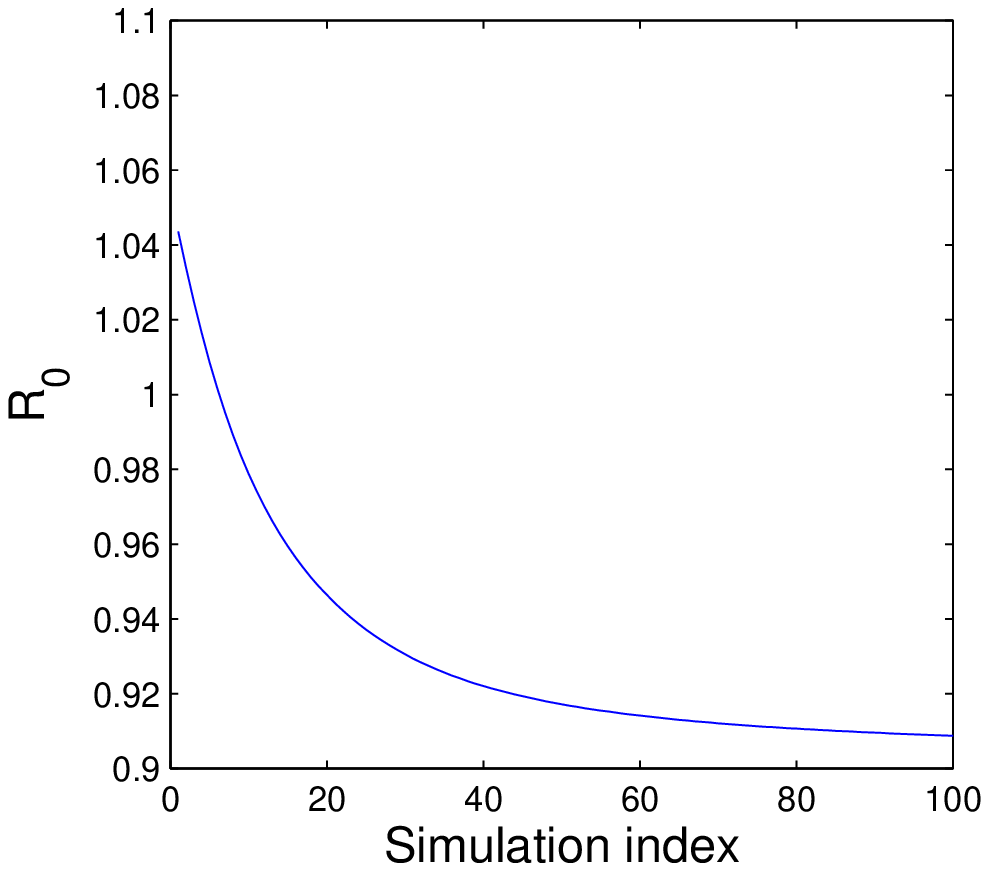}}
\hspace{0.1in}
\subfigure[As the livestock movement rate  from node $i$ to node $j$ ($\omega_{2ij}$) increases when $\gamma_{2i}>\gamma_{2j} $, $R_0$ increases.]{
\label{fig:moverecoverlsol}
\includegraphics[angle=0,width=6cm,height=5cm]{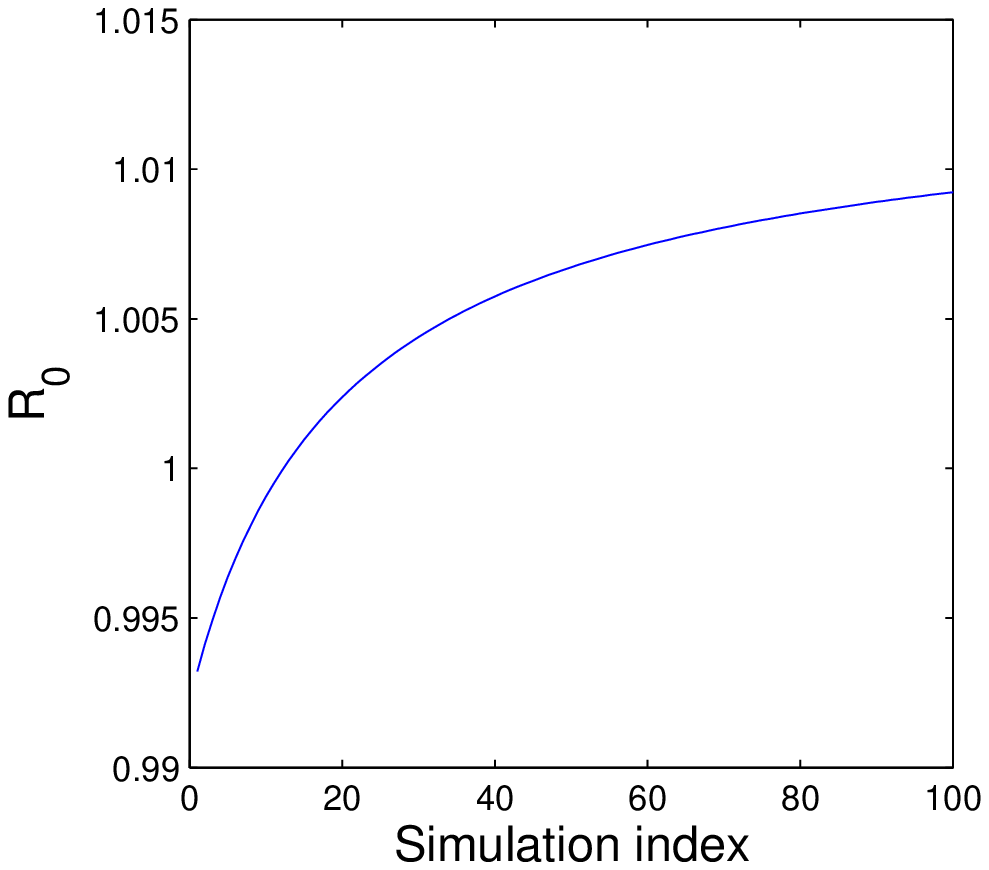}}
\hspace{0.1in}
\caption{The reproduction number for  four-node networks   with different livestock recovery  rates during one hundred  runs. }
\label{fig:differentrecoveryrates}
\end{figure}
\begin{figure}[!htbp]
\centering
\subfigure[As the  livestock movement rate from node $j$ to node $i$ ($\omega_{2ji}$) increases  when $\mu_{2i}>\mu_{2j} $, $R_0$ decreases.]{
\label{fig:movemortalityltos}
\includegraphics[angle=0,width=6cm,height=5cm]{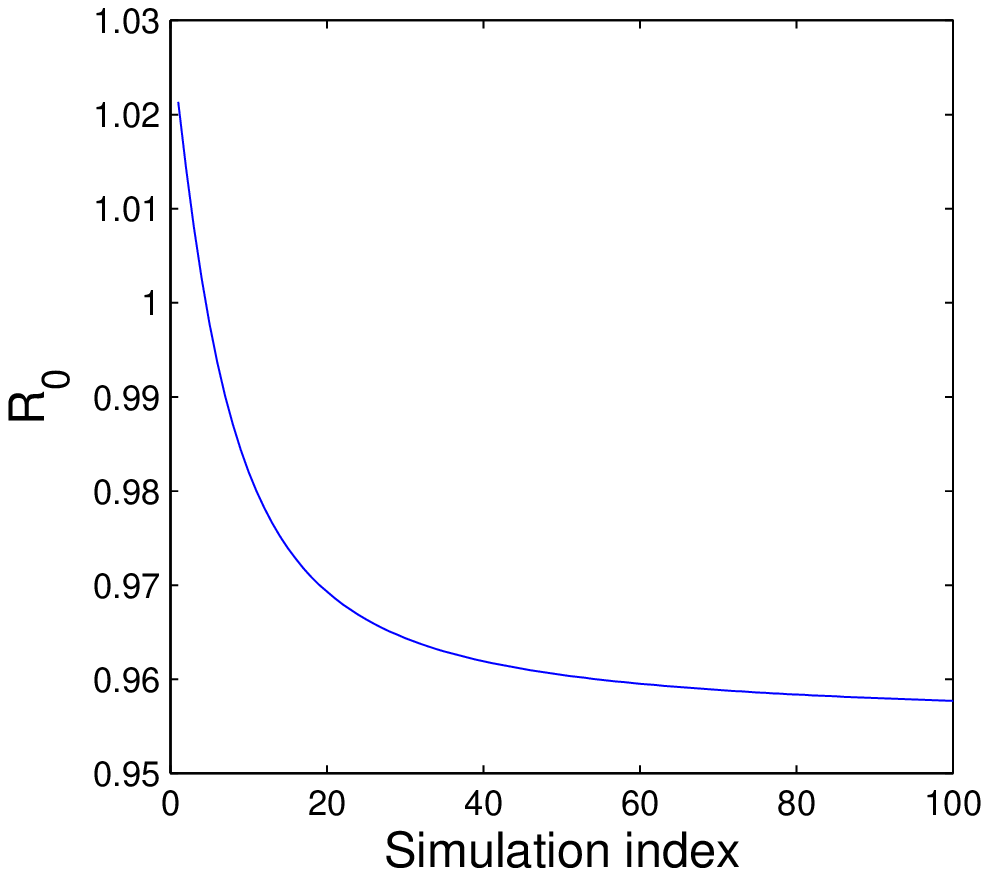}}
\hspace{0.1in}
\subfigure[As the  livestock movement rate  from node $i$ to node $j$ ($\omega_{2ij}$) increases  when $\mu_{2i}>\mu_{2j} $,  $R_0$ increases.]{
\label{fig:movesmortalitysol}
\includegraphics[angle=0,width=6cm,height=5cm]{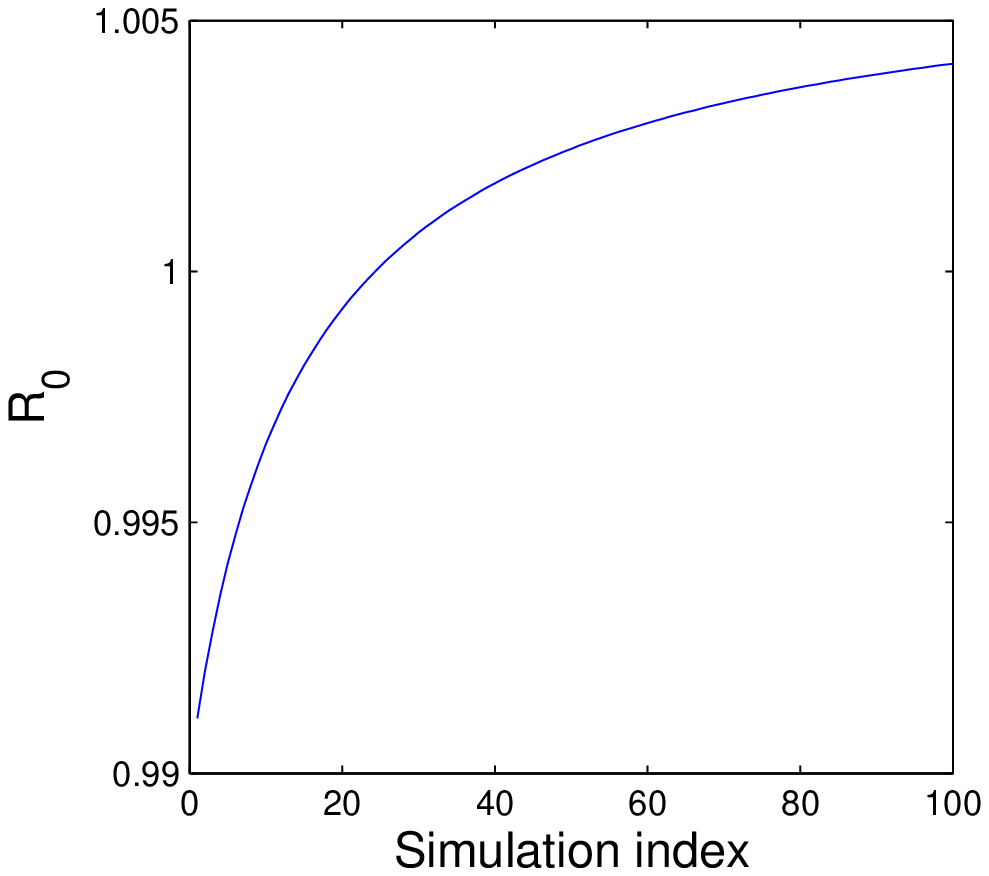}}
\hspace{0.1in}
\caption{The reproduction number for  four-node networks   with different livestock mortality  rates during one hundred  runs.}
\label{fig:differentmortalityrates}
\end{figure}
\end{small}

\section{Results and discussions}
\label{section:result}
We propose an explicit expression of  $R_0$, which  is formulated as a function of vertical and horizontal transmission parameters shown in Equation $(\ref{equation:R0general})$. This formula facilitates computing $R_0$  for many diseases that involve both vertical and horizontal transmission by  replacing the spectral radius  of  the original next generation matrix with that of a smaller size matrix.  The  lower bound of $R_0$  equals the reproduction number for horizontal transmission.     We applied Equation $(\ref{equation:R0general})$  to the RVF model, deriving   $R_0$  and its lower and  upper  bounds. We compared  the tightness of different bounds, and  analyzed the role of livestock movement rates and disease parameters on $R_0$  through numerical simulations.

  The reproduction number  for RVF meta-population model relates to  the reproduction number for horizontal transmission, involving \it Aedes\rm-livestock  interaction and \it Culex\rm-livestock interaction, and vertical transmission parameters.      Different bounds of $R_0$ for heterogeneous networks are given by  Theorem  \ref{theorem:R0}, Theorem \ref{theorem:FHVHinverse},  and Theorem \ref {theorem:heterogeneousparameter} with decreasing tightness and increasing easiness. 
For homogeneous networks, the reproduction number for horizontal transmission in Equation $(\ref{equation:R0HRVF})$ and  bounds of $R_0$ given by Corollary $\ref{cor:homogeneous}$ are proved  independent of livestock movement rates,  and    equal to corresponding terms for homogeneous  populations presented in \cite {LingXue2010}.   The lower bound is  the reproduction  number for horizontal transmission and upper bound is the sum of the reproduction number for  horizontal transmission and the largest  transovarial transmission rate of \it Aedes \rm mosquitoes among nodes.

Typically  networks in the real world are heterogeneous. Rates of livestock death, incubation, mortality, recovery, and contact with mosquitoes can vary in different nodes due to climate, public health facilities, environment, and/or type of nodes (e.g., death rates of livestock in feedlots are higher than those in livestock premises). Variations in weather may affect values of some mosquito parameters, e.g.,   rainfall affects mosquito birth rates, and  temperature affects mosquito  incubation rates.  Even if weather conditions are homogeneous across all nodes, different genera and/or species of mosquitoes can exhibit different rates of incubation, contact, death, birth, and/or birth.   Numerical  simulations  show livestock movement rates between different nodes only affect $R_0$ when the network is spatially heterogeneous regarding parameters. Changing livestock movement rates    on heterogeneous networks results in $R_0$ varying  between  values below and above the critical value $1$.  When other parameters remain homogeneous and constant, increasing livestock movement rates from nodes with smaller contact rates to those with larger contact rates increases $R_0$.  If livestock movement rates are increased from nodes with smaller livestock death rates (or recovery rates, or mortality rates) to nodes with larger livestock death rates (or recovery rates, or mortality rates), $R_0$ decreases. This observation helps us better envision effective mitigation strategies executing movement bans between some nodes and in some directions.

Whatever heterogeneity exists between nodes, our same mathematical model in Equations $(\ref{equation:generalP})$  through $(\ref{equation:generalR})$, and the explicit expression of $R_0$ in $(\ref{equation:R0general})$, are applicable.  Our formula for $R_0$ presented in this paper can be used for numerous diseases models aside from RVF.

Our work on RVF contributes computing  $R_0$ accurately by taking into account vertical transmission, which is  important but ignored by  modelers.  We simplified the derivation of $R_0$ by computing the spectral radius of a smaller size matrix than the original next generation matrix.  Bounds of $R_0$ facilitate estimating $R_0$ of RVF metapopulation model. The simulation results on livestock movement rates and parameters are helpful in developing  efficient mitigation strategies for RVF.

\section*{Acknowledgments}
This work has been supported by the DHS Center of Excellence for Emerging and Zoonotic Animal Diseases (CEEZAD). We are grateful to the effort made by the anonymous reviewers. We would like to give thanks to Alice Trussell and Andrea Engelken for help on English proofreading.

\bibliographystyle{model1-num-names}
\bibliography{networkR0}

\section*{Appendix}

{\bf Proof of Theorem \ref{theorem:R0}.}\quad
 The left inequality is the same as (\ref{eq:roandroh}). We now show that the right inequality holds.
 By  (\ref{equation:U})  and (\ref{equation:VHinverse}),
\begin{equation*}
-W( \oplus _{i=1}^{n}\theta_{1i}^{-1})UV_H^{-1}= \left[ {%
\begin{array}{cccccccccccccccc}
 0_{4n\times 4n}&  0_{4n \times 4n} \\
   Y & Z\\
\end{array}
} \right],\ {\rm where}
\end{equation*}
\begin{equation*}
Y= \left[ {%
\begin{array}{cccccccccccccccc}
X_{1}^{-1}(\oplus _{i=1}^{n}(b_{1i}q_{1i} \varepsilon_{1i}))M_{1}^{-1}&  0_{n \times 3n}\\
   0_{3n\times n}  & 0_{3n \times 3n}\\
\end{array}
} \right],\ \
Z= \left[ {%
\begin{array}{cccccccccccccccc}
(\oplus _{i=1}^{n}(b_{1i}q_{1i}))X_{1}^{-1}&  0_{n \times 3n} \\
    0_{3n\times n}  & 0_{3n \times 3n}\\
\end{array}
} \right].
\end{equation*}
Note that $X_1$ and   $X_1^{-1}$ are diagonal matrices. Moreover, the nonzero eigenvalues of $- W(\oplus _{i=1}^{n}\theta_{1i}^{-1}) UV_H^{-1}$ are diagonal entries of $(\oplus _{i=1}^{n}(b_{1i}q_{1i})) X_1^{-1}$. Hence,  $- W(\oplus _{i=1}^{n}\theta_{1i}^{-1})UV_H^{-1}=\mathcal{P}\mathcal{D}\mathcal{P}^{-1}$ for some $\mathcal{P}$. Here
  \begin{equation*}
\mathcal{D}
=\left[ {%
\begin{array}{cccccccccccccccc}
0_{4n \times 4n}&0_{4n \times 4n}\\
 0_{4n \times 4n}& \mathcal{Q} \\
\end{array}
} \right],\quad
\mathcal{Q}
=\left[ {%
\begin{array}{cccccccccccccccc}
0_{3n \times 3n} &  0_{3n \times n} \\
0_{n \times 3n}& (\oplus _{i=1}^{n}(b_{1i}q_{1i}))X_{1}^{-1} \\
\end{array}
} \right].
\end{equation*}
%
%
%
From linear algebra, each column of  $\mathcal{P}$ can be chosen as an eigenvector of $-W(\oplus _{i=1}^{n}\theta_{1i}^{-1})UV_H^{-1}$. By direct calculation,
\begin{align*}
\mathcal{P}
=\left[ {%
\begin{array}{cccccccccccccccc}\mathcal{H}_{4n \times 4n}&0\\
\mathcal{J}_{4n \times 4n}&  \mathcal{L}_{4n \times 4n}\\
\end{array}
} \right],\quad {\rm where}\quad
\mathcal{H}
=\left[ {
\begin{array}{cccccccccccccccc}(\oplus _{i=1}^{n}(b_{1i}q_{1i}))X_{1}^{-1}&0_{n \times 3n} \\
0_{3n \times n}& I_{3n\times 3n}\\
\end{array}
} \right],\\
\mathcal{L}
=\left[ {
\begin{array}{cccccccccccccccc}0_{n\times 3n} &I_{n \times n} \\
I_{3n \times 3n}& 0_{3n \times n}
\end{array}
} \right],\quad \mathcal{J}
=\left[ {
\begin{array}{cccccccccccccccc}- (\oplus _{i=1}^{n}(b_{1i}q_{1i}\varepsilon_{1i}))X_{1}^{-1}M_{1}^{-1} &0_{n \times 3n} \\
0_{3n\times n}& 0_{3n\times 3n}\\
\end{array}
} \right].
\end{align*}
Since
$F_HV_H^{-1}-W(\oplus _{i=1}^{n}\theta_{1i}^{-1})UV_H^{-1}=\mathcal{P}(\mathcal{P}^{-1}F_HV_H^{-1}\mathcal{P}+\mathcal{D})\mathcal{P}^{-1}$, we have
\begin{equation}\label{equation:FVinverse}
\rho(FV^{-1})= \rho(F_HV_H^{-1}-W(\oplus _{i=1}^{n}\theta_{1i}^{-1})UV_H^{-1}) =  \rho(\mathcal{P}^{-1}F_HV_H^{-1}\mathcal{P}+\mathcal{D}) .
\end{equation}
We  clam that $\mathcal{P}^{-1}F_HV_H^{-1}\mathcal{P}$ is a nonnegative matrix. By calculation,
 $$\mathcal{P}^{-1}
=\left[ {
\begin{array}{cccccccccccccccc}\mathcal{H}^{-1}&0\\
-\mathcal{L}^{-1}\mathcal{J}\mathcal{H}^{-1}&  \mathcal{L}^{-1}
\end{array}
} \right],\
\mathcal{H}^{-1}
=\left[ {
\begin{array}{cccccccccccccccc}(\oplus _{i=1}^{n}\frac{1}{b_{1i}q_{1i}})X_1 &0_{n \times 3n} \\
0_{3n \times n}& I_{3n \times 3n}
\end{array}
} \right],\ \
\mathcal{L}^{-1}
=\left[ {
\begin{array}{cccccccccccccccc}0_{3n \times n} &I_{3n \times 3n} \\
I_{n \times n}& 0_{n \times 3n}
\end{array}
} \right].$$
It is clear that  $\mathcal{H}^{-1}$, $\mathcal{L}^{-1}$, and $- \mathcal{L}^{-1}\mathcal{J}\mathcal{H}^{-1} $ are all nonnegative matrices. Hence, $\mathcal{P}^{-1}$ is a nonnegative matrix. We now show that $F_HV_H^{-1}\mathcal{P}$ is a nonnegative matrix.
\begin{equation*}
F_HV_H^{-1}\mathcal{P}
=\left[ {
\begin{array}{cccccccccccccccc}
 \mathcal{A} (\oplus _{k=1}^{4}\mathcal{Z}_k)\mathcal{H}+\mathcal{A}(\oplus _{k=1}^{4}X_k^{-1})\mathcal{J}&\mathcal{A}(\oplus _{k=1}^{4}X_k^{-1})\mathcal{L}\\
0& 0\\
\end{array}
} \right],
\end{equation*}
where $\mathcal{A}(\oplus _{k=1}^{4}X_k^{-1})\mathcal{L}$ is a nonnegative matrix and $\mathcal{Z}_k=X_k^{-1}(\oplus _{i=1}^{n}\varepsilon_{ki}) M_k^{-1}$.
Furthermore, the only possible negative entries of $\mathcal{A}(\oplus _{k=1}^{4}\mathcal{Z}_k)\mathcal{H}+\mathcal{A}(\oplus _{k=1}^{4}X_k^{-1})\mathcal{J}$ are in its $(2,1)$ and $(4,1)$ blocks. But the block in $(2,1)$-entry is
\begin{equation*}
\mathcal{A}_{21}X_1^{-1}( \oplus _{i=1}^{n}\varepsilon_{1i})M_1^{-1}  ( \oplus _{i=1}^{n}(b_{1i}q_{1i}))X_1^{-1}+\mathcal{A}_{21}X_1^{-1} (-\oplus _{i=1}^{n}(b_{1i}q_{1i}\varepsilon_{1i}))X_1^{-1}M_1^{-1}=0.
\end{equation*}
By assumption, $X_1$ and $M_1$ are both diagonal matrices. The last equality follows $X_1^{-1}M_1^{-1}=M_1^{-1}X_1^{-1}$. Similarly, the block in $(4,1)$-entry is
\begin{equation*}
\mathcal{A}_{41}X_1^{-1}( \oplus _{i=1}^{n}\varepsilon_{1i})M_1^{-1}  ( \oplus _{i=1}^{n}(b_{1i}q_{1i}))X_1^{-1}+\mathcal{A}_{41}X_1^{-1} (-\oplus _{i=1}^{n}(b_{1i}q_{1i}\varepsilon_{1i}))X_1^{-1}M_1^{-1}=0.
\end{equation*}
Hence, $F_HV_H^{-1}\mathcal{P}$ is a nonnegative matrix. This proves the claim.
By Theorem $2$ in \cite{Cohen1979}, we have
\begin{equation}\label{equation:upperboundofR0}
\rho(FV^{-1})\leq \rho(\mathcal{P}^{-1}F_HV_H^{-1}\mathcal{P})+\rho(\mathcal{D})= \rho(F_HV_H^{-1})+\rho(\mathcal{D}).
\end{equation}
Since $X_1=\oplus_{i=1}^n\frac{d_{1i}N_{1i}^0}{K_1}$ and $N^0_{1i}=\frac{b_{1i}K_1}{d_{1i}}$, we further have
\begin{eqnarray*}
\rho(\mathcal{D})
=\rho(-W(\oplus _{i=1}^{n}\theta_{1i}^{-1})UV_H^{-1})=\rho((\oplus _{i=1}^{n}(b_{1i}q_{1i}))X_1^{-1})
=\rho(\oplus _{i=1}^{n}q_{1i})\leqslant \max_i(q_{1i}).
\end{eqnarray*}
Therefore,
\begin{equation*}
\rho(F_HV_H^{-1}) \leq  R_0=\rho(FV^{-1})\leq  \rho(F_HV_H^{-1})+\max_i(q_{1i}). \label{equation:upperboundofR0}
\end{equation*}
 This finishes the proof. \vspace{10pt}


{\bf Proof of Theorem \ref{theorem:FHVHinverse}.}\quad
By  Equations  $(\ref{equation:VHinverse})$ and $(\ref{equation:FH})$,
\begin{align*}
F_HV_H^{-1}= \left[ {%
\begin{array}{cccccccccccccccc}
 \mathcal{A} (\oplus _{k=1}^{4} \mathcal{Z}_k)&\mathcal{A}(\oplus_{k=1}^{4} X_k^{-1})  \\
 0& 0\\
\end{array}
} \right].
\end{align*}
Then
\begin{equation}
R_0^H=\rho(F_HV_H^{-1})=\rho(\mathcal{A} (\oplus_{k=1}^{4}\mathcal{Z}_k)). \label{equation: FHVHinverse}
\end{equation}
 By Equation $(\ref{equation:mathcalA})$,
\begin{align*}
\mathcal{A}( \oplus_{k=1}^{4} \mathcal{Z}_k)&= \left[ {%
\begin{array}{cccccccccccccccc}
  0&\mathcal{A}_{12}\mathcal{Z}_2&0&0\\
\mathcal{A}_{21}\mathcal{Z}_1& 0&\mathcal{A}_{23}\mathcal{Z}_3&0\\
 0& \mathcal{A}_{32}\mathcal{Z}_2&0&0\\
  \mathcal{A}_{41}\mathcal{Z}_1& \mathcal{A}_{42}\mathcal{Z}_2& \mathcal{A}_{43}\mathcal{Z}_3&0
\end{array}
} \right]
=: \left[ {%
\begin{array}{cccccccccccccccc}
  0&\mathcal{B}_1&0&0\\
 \mathcal{B}_2& 0& \mathcal{B}_3&0\\
 0& \mathcal{B}_4&0&0\\
\mathcal{B}_5& \mathcal{B}_6&\mathcal{B}_7&0\\
\end{array}
} \right].
\end{align*}

To compute the eigenvalues of   $\mathcal{A}( \oplus_{k=1}^{4} \mathcal{Z}_k)$, we first calculate the characteristic polynomial of $\mathcal{A}( \oplus_{k=1}^{4} \mathcal{Z}_k)$ as follows.
\begin{align*}
&|\lambda I_{4n}-\mathcal{A}( \oplus_{k=1}^{4} \mathcal{Z}_k)|= \left| {%
\begin{array}{cccccccccccccccc}
  \lambda  I_n & -\mathcal{B}_1&0&0\\
  -\mathcal{B}_2&  \lambda  I_n&  -\mathcal{B}_3&0\\
 0& - \mathcal{B}_4&  \lambda  I_n&0\\
 - \mathcal{B}_5& - \mathcal{B}_6&  - \mathcal{B}_7&\lambda  I_n\\
\end{array}
} \right|= \lambda^n\left| {%
\begin{array}{cccccccccccccccc}
  \lambda  I_n & -\mathcal{B}_1&0\\
  -\mathcal{B}_2&  \lambda  I_n&  -\mathcal{B}_3\\
 0& - \mathcal{B}_4&  \lambda  I_n\\
\end{array}
} \right| \vspace{10pt}\\
&= \lambda^n\Big|\left[ {%
\begin{array}{cccccccccccccccc}
  I_n&\lambda \mathcal{B}_2^{-1}&0\\
0& I_n& 0\\
 0&0&I_n\\
\end{array}
} \right]\left[ {%
\begin{array}{cccccccccccccccc}
  \lambda  I_n & -\mathcal{B}_1&0\\
  -\mathcal{B}_2&  \lambda  I_n&  -\mathcal{B}_3\\
 0& - \mathcal{B}_4&  \lambda  I_n\\
\end{array}
} \right]\Big|
= \lambda^n\Big|\left[ {%
\begin{array}{cccccccccccccccc}
  0&  -\mathcal{B}_1+\lambda^2 \mathcal{B}_2^{-1}&-\lambda \mathcal{B}_2^{-1}\mathcal{B}_3\\
  -\mathcal{B}_2& \lambda I_n&   -\mathcal{B}_3\\
 0&  -\mathcal{B}_4&\lambda I_n\\
\end{array}
} \right]\Big| \vspace{10pt}\\
&= \lambda^n|\mathcal{B}_2|\left| {%
\begin{array}{cccccccccccccccc}
-\mathcal{B}_1+\lambda^2 \mathcal{B}_2^{-1}&-\lambda \mathcal{B}_2^{-1}\mathcal{B}_3\\
-\mathcal{B}_4& \lambda I_n\\
\end{array}
} \right|
= \lambda^n|\mathcal{B}_2|\left| \Big[{%
\begin{array}{cccccccccccccccc}
-\mathcal{B}_1+\lambda^2 \mathcal{B}_2^{-1}&-\lambda \mathcal{B}_2^{-1}\mathcal{B}_3\\
-\mathcal{B}_4& \lambda I_n\\
\end{array}} \right]\left[ {%
\begin{array}{cccccccccccccccc}
  I_n&\lambda \mathcal{B}_4^{-1}\\
 0& I_n\\
\end{array}
} \right]\Big|\vspace{10pt}\\
&= \lambda^n|\mathcal{B}_2|\left| {%
\begin{array}{cccccccccccccccc}
-\mathcal{B}_1+\lambda^2 \mathcal{B}_2^{-1}&-\lambda(\mathcal{B}_1\mathcal{B}_4^{-1}-\lambda^2\mathcal{B}_2^{-1}\mathcal{B}_4^{-1}+\mathcal{B}_2^{-1}\mathcal{B}_3)\\
-\mathcal{B}_4& 0\\
\end{array}
} \right|\vspace{10pt}\\
&=\lambda^n |\mathcal{B}_2|\left| {%
\begin{array}{cccccccccccccccc}
-\lambda(\mathcal{B}_1\mathcal{B}_4^{-1}-\lambda^2\mathcal{B}_2^{-1}\mathcal{B}_4+\mathcal{B}_2^{-1}\mathcal{B}_3)&\mathcal{B}_1-\lambda^2 \mathcal{B}_2^{-1}\\
 0&\mathcal{B}_4\\
\end{array}
} \right|\vspace{10pt}\\
&=\lambda^{n} |\mathcal{B}_2||\mathcal{B}_4|\left| {%
\begin{array}{cccccccccccccccc}
-\lambda (\mathcal{B}_1\mathcal{B}_4^{-1}-\lambda^2\mathcal{B}_2^{-1}\mathcal{B}_4^{-1}+\mathcal{B}_2^{-1}\mathcal{B}_3)\\
\end{array}
} \right|\vspace{10pt}\\
&= \lambda^{2n}|\mathcal{B}_2||\mathcal{B}_4|\left| {%
\begin{array}{cccccccccccccccc}
\lambda^2 \mathcal{B}_2^{-1}\mathcal{B}_4^{-1}-(\mathcal{B}_1\mathcal{B}_4^{-1}+\mathcal{B}_2^{-1}\mathcal{B}_3)\\
\end{array}
} \right|\vspace{10pt}\\
&= \lambda^{2n}|\mathcal{B}_2||\mathcal{B}_4||\mathcal{B}_2^{-1}||\mathcal{B}_4^{-1}|\left| {%
\begin{array}{cccccccccccccccc}
\lambda^2I_n -(\mathcal{B}_4\mathcal{B}_2\mathcal{B}_1\mathcal{B}_4^{-1}+\mathcal{B}_4\mathcal{B}_3)\\
\end{array}
} \right|\vspace{10pt}\\
&= \lambda^{2n}\left| {%
\begin{array}{cccccccccccccccc}
\lambda^2I_n -(\mathcal{B}_4\mathcal{B}_2\mathcal{B}_1\mathcal{B}_4^{-1}+\mathcal{B}_4\mathcal{B}_3)\\
\end{array}
} \right|.
\end{align*}

Matrix $\mathcal{A}( \oplus_{k=1}^{4} \mathcal{Z}_k)$ has $2n$ zero eigenvalues. The spectral radius of $\mathcal{A}( \oplus_{k=1}^{4} \mathcal{Z}_k)$ is the square root of the spectral  radius of $\mathcal{B}_4\mathcal{B}_2\mathcal{B}_1\mathcal{B}_4^{-1}+\mathcal{B}_4\mathcal{B}_3$.  By  Equation $(\ref{equation: FHVHinverse})$, we obtain
\begin{align}
\rho(F_HV_H^{-1})=\sqrt{\rho(\mathcal{B}_4\mathcal{B}_2\mathcal{B}_1\mathcal{B}_4^{-1}+\mathcal{B}_4\mathcal{B}_3)}
=\sqrt{\rho(\mathcal{B}_4(\mathcal{B}_2\mathcal{B}_1+\mathcal{B}_3\mathcal{B}_4)\mathcal{B}_4^{-1}})
=\sqrt{\rho(\mathcal{B}_2\mathcal{B}_1+\mathcal{B}_3\mathcal{B}_4)}. \label{equation:rhoRHVHinverse}
\end{align}
 Recall that $\mathcal{A}_{21}$, $\mathcal{A}_{12}$,  $X_1$, $M_1$,  $\mathcal{A}_{23}$, $\mathcal{A}_{32}$,  $M_3$, $X_3$ are all diagonal matrices. By  the assumption that $\varepsilon_{2i}=\varepsilon_{2}$, for all $i $, we obtain
\begin{align*}
\mathcal{B}_2\mathcal{B}_1&=( \oplus _{i=1}^{n}\varepsilon_{1i}\varepsilon_{2})\mathcal{A}_{21}X_1^{-1}M_1^{-1}\mathcal{A}_{12}X_2^{-1}M_2^{-1}=(\oplus _{i=1}^{n}\frac{\varepsilon_{2}\varepsilon_{1i}\beta_{12i}\beta_{21i}}{b_{1i}(b_1+\varepsilon_{1i})})X_2^{-1}M_2^{-1},\\
\mathcal{B}_3\mathcal{B}_4&=( \oplus _{i=1}^{n}\varepsilon_{2}\varepsilon_{3i})\mathcal{A}_{23}X_3^{-1}M_3^{-1}\mathcal{A}_{32}X_2^{-1}M_2^{-1}=(\oplus _{i=1}^{n}\frac{\varepsilon_{2}\varepsilon_{3i}\beta_{32i}\beta_{23i}}{b_{3i}(b_{3i}+\varepsilon_{3i})})X_2^{-1}M_2^{-1}.
\end{align*}
By the definition of $\chi_i$ in (\ref{chii}), we have
\begin{align*}
\min_i(\chi_i)\rho(X_2^{-1}M_2^{-1}) \leqslant \rho(\mathcal{B}_2\mathcal{B}_1+\mathcal{B}_3\mathcal{B}_4) \leqslant \max_i(\chi_i)\rho(X_2^{-1}M_2^{-1}).
\end{align*}
Therefore,
\begin{align*}
\sqrt{\min_i(\chi_i)    \rho(X_2^{-1}M_2^{-1})} \leqslant  \rho(F_HV_H^{-1})  \leqslant \sqrt{\max_i(\chi_i)  \rho(X_2^{-1}M_2^{-1})}.
\end{align*}
According to Theorem \ref{theorem:R0},
\begin{align*}
\sqrt{\min_i(\chi_i)    \rho(X_2^{-1}M_2^{-1})} \leqslant  R_0 \leqslant \sqrt{\max_i(\chi_i)  \rho(X_2^{-1}M_2^{-1})}+\max_i(q_{1i}).
\end{align*} This finishes the proof. \vspace{10pt}


{\bf Proof of Corollary \ref{cor1}.}
By the assumptions in (\ref{parameter2}), we have ${\rm min}_i (\chi_i)=\chi={\rm max}_i(\chi_i)$. Corollary follows from Theorem \ref{theorem:FHVHinverse}.\\

{\bf Proof of Theorem \ref{theorem:heterogeneousparameter}.}\quad
According to Theorem $\ref{theorem:column}$,
 \begin{align}
\frac{1}{\max_i{(d_{2i}+\varepsilon_{2i})\max_i(d_{2i}+\gamma_{2i}+\mu_{2i})}}   \leqslant  \rho(X_2^{-1}M_2^{-1})   \leqslant  \frac{1}{\min_i{(d_{2i}+\varepsilon_{2i})\min_i(d_{2i}+\gamma_{2i}+\mu_{2i})}}.
\end{align}
By Theorem \ref{theorem:FHVHinverse}, we have
 \begin{align*}
\sqrt{ \frac{\min_i{(\chi_i)}}{\max_i{(d_{2i}+\varepsilon_{2i})\max_i(d_{2i}+\gamma_{2i}+\mu_{2i})}}}  \leqslant  R_0 \leqslant \sqrt{ \frac{\max_i{(\chi_i)}}{\min_i({d_{2i}+\varepsilon_{2i})\min_(d_{2i}+\gamma_{2i}+\mu_{2i})}}}+\max_i(q_{1i}).
\end{align*}
This finishes the proof. \vspace{10pt}

{\bf Proof of Corollary \ref{cor:homogeneous}.}
 By the conditions in $(\ref{parameter2})$ and $(\ref{parameter3})$, we have
\begin{align*}
  \frac{\min_i{(\chi_i)}}{\max_i{(d_{2i}+\varepsilon_{2i})\max_i(d_{2i}+\gamma_{2i}+\mu_{2i})}}=\frac{\chi  }{(d_2+\varepsilon_2)(d_2+\gamma_2+\mu_2)}= \frac{\max_i{(\chi_i)}}{\min_i({d_{2i}+\varepsilon_{2i})\min_(d_{2i}+\gamma_{2i}+\mu_{2i})}}.
\end{align*}
Corollary follows from Theorem \ref{theorem:heterogeneousparameter}.\vspace{10pt}
\begin{thm}\label{theorem:spectral}
  If both $A$ and $B$ are non-negative square matrices, then
  $\rho(A) \leq \rho(A+B).$
\end{thm}
{\bf Proof.}
 Recall that  the Gelfand's formula is that
$$\rho(A)=\lim_{k\rightarrow \infty}\|A^k\|^{\frac{1}{k}}$$ for any matrix norm $\|\cdot \|$.
If $A, B$ are both non-negative, then $A\leq A+B$. Hence, $0\leq A^k \leq (A+B)^k$ for any $k\in \mathbb{N}$. By the property of matrix norm,
$0 \leq \|A^k\| \leq \|(A+B)^k\|$. Thus,
$$0\leq \lim_{k\rightarrow \infty}\|A^k\|^{\frac{1}{k}} \leq \lim_{k\rightarrow \infty}\|(A+B)^k\|^{\frac{1}{k}}.$$
The theorem follows from the Gelfand's formula.

\begin{thm} \label{theorem:N2}
 For the model presented  in Section $\ref{subsection:general model}$  (Equations $(\ref{equation:generalP})$ through $(\ref{equation:generalR})$), a unique nonnegative  solution for total number of  species $k$  individuals  in node $i$ at  DFE exists.
\end{thm}
{\bf Proof.}
To solve the total number of species $k$  individuals in each node at DFE, we need to solve the following system of equations.
\begin{equation}
\mathcal{W} \begin{bmatrix}
N_{k1}^*&
N_{k2}^*&
\cdots  &
N_{kn}^*
\end{bmatrix}^T
= \begin{bmatrix}
r_{k1}&
r_{k2}&
\cdots &
r_{kn}
\end{bmatrix}^T, \label{equation:DFE}
\end{equation}
where
\begin{equation*}
 \mathcal{W} =\left[ {%
\begin{array}{cccccccccccccccc}
d_{k1}+\sum_{j=2}^n\omega_{k1j}& -\omega_{k21} &  \cdots &   -\omega_{kn1}  \\
 -\omega_{k12} & d_{k2}+\sum_{j=1, j \neq 2}^n\omega_{k2j}&  \cdots&  -\omega_{kn2}  \\
\cdots & \cdots& \cdots &\cdots \\
 -\omega_{k1n} & -\omega_{k2n}& \cdots &d_{kn}+\sum_{j=1}^{n-1}\omega_{knj}\\
\end{array}
} \right].
\end{equation*}
The variable vector
$\left[{%
\begin{array}{cccccccccccccccc}
N_{k1}^* &  N_{k2}^* \cdots N_{kn}^*
\end{array}
} \right] ^T$ is   to be solved.
We note that $ \mathcal{W} $ is a diagonal dominant matrix of its column entries \cite{cheng2005}, i.e.,   $ \mathcal{W}_{ii} \geqslant \sum_{i=1, i \neq j}^n  \mathcal{W}_{ij}  $,   for all $i$, where $ \mathcal{W}_{ij}$ is the $(i,j)$ entry of $ \mathcal{W} $. By Theorem $1$ in page $654$ of  \cite{cheng2005}, $ \mathcal{W} $ is  invertible.   Moreover,  by Theorem $\ref{theorem:X2M2}$  in  appendix, $ \mathcal{W}^{-1} $ is nonnegative.  Thus,  there exists a unique nonnegative solution for the system of  equations  $(\ref{equation:DFE})$.The unique nonnegative solution is
 $$\left[ {%
\begin{array}{cccccccccccccccc}
N_{k1}^* & N_{k2}^* &  \cdots  & N_{kn}^*\\
\end{array}
} \right]^T= \left[ {%
\begin{array}{cccccccccccccccc}
N_{k1}^0 & N_{k2}^0 &  \cdots  & N_{kn}^0\\
\end{array}
} \right]^T=   \mathcal{W}^{-1}\left[ {%
\begin{array}{cccccccccccccccc}
r_{k1} &r_{k2}& \cdots& r_{kn} \\
\end{array}
} \right]^T.$$

\begin{thm}  \label{theorem:column}
Let $A_k \ (k=1, 2, \cdots , m)$ be an $n \times n$ diagonal dominant matrix with $A_k^{-1} \geq 0$. Denote the $(i, j)  $ entry of $A_k$ by $a_{kij}$. Let  $a_k^L=\min_j(\sum_i a_{kij})>0$, $a_k^H=\max_j(\sum_i a_{kij})$, where $j=1, 2, \cdots, n$, then $$\prod_{k=1}^m \frac{1}{a_k^H}\leqslant  \rho(\prod_{k=1}^mA_k^{-1}) \leqslant \prod_{k=1}^m \frac{1}{a_k^L}.$$
\end{thm}
{\bf Proof.}
Clearly,
$0\leq a_k^L\mathcal{C}  \leqslant   \mathcal{C} A_k \leqslant a_k^H\mathcal{C},$
where
$\mathcal{C} =[1 \ 1  \cdots \ 1 ]_{1 \times n}.$
Since $A_k^{-1}\geqslant 0$, we obtain
$$0 \leq \frac{\mathcal{C}}{a_k^H} \leq \mathcal{C}A_k^{-1} \leq  \frac{\mathcal{C}}{a_k^L}.$$
Similarly, $$0  \leq \frac{\mathcal{C}}{a_k^Ha_{k-1}^H} \leq \frac{\mathcal{C}A_{k-1}^{-1}}{a_k^H} \leq \mathcal{C}A_k^{-1}A_{k-1}^{-1} \leq  \frac{\mathcal{C}A_{k-1}^{-1}}{a_k^L}\leq \frac{\mathcal{C}}{a_k^La_{k-1}^L}.$$
Following the same argument,
\begin{equation*}
\prod_{k=1}^m \frac{1}{a_k^H}\mathcal{C}\leqslant  \mathcal{C}\prod_{k=1}^m A_k^{-1}\leqslant \prod_{k=1}^m \frac{1}{a_k^L} \mathcal{C}.
\end{equation*}
By Corollary $1$ in \cite{Cohen1979}, any $n \times n$  nonnegative matrix $A$  satisfies: \\
      \begin{equation}
\min_j(\sum_{i=1}^na_{ij})\leqslant  \rho(A) \leqslant \max_j(\sum_{i=1}^na_{ij}).  \label{equation:rhoA}
\end{equation}
Because the entries of $\mathcal{C}\prod_{k=1}^m A_k^{-1}$ is the sum of each column of matrix $\prod_{k=1}^m A_k^{-1}$, by Inequality $(\ref{equation:rhoA})$,
\begin{equation*}
\prod_{k=1}^m \frac{1}{a_k^H}\leqslant  \rho(\prod_{k=1}^mA_k^{-1}) \leqslant \prod_{k=1}^m \frac{1}{a_k^L}.
\end{equation*}

\begin{thm} \label{theorem:X2M2}
Matrices  $M_k$ and $X_k$  in  $( \ref{matrix:generalMk})$  
 are invertible, and $M_k$ and $X_k$ are nonnegative.
Moreover,  Matrices $M_k^{-1}$ and  $X_k^{-1}$ are nonnegative matrices.
\end{thm}
{\bf Proof.}
Note that $M_k $ is a diagonal dominant matrix of its column entries. By Theorem $1$ in page $654$ of  \cite{cheng2005}, $M_k$ and $X_k$ are invertible.
We now prove that $M_k^{-1}$  is nonnegative. Matrix $M_k$ can be rewritten as follows.
\begin{eqnarray*}
M_k= \left[ {%
\begin{array}{cccccccccccccccc}
\zeta_{k1}& -\omega_{k21} &  \cdots &   -\omega_{kn1}  \\
 -\omega_{k12} & \zeta_{k2}&  \cdots&  -\omega_{kn2}  \\
\cdots & \cdots& \cdots &\cdots \\
 -\omega_{k1n} & -\omega_{k2n} & \cdots &\zeta_{kn}\\
\end{array}
} \right] 
= \oplus_{i=1}^n\zeta_{ki}
- \left[ {%
\begin{array}{cccccccccccccccc}
0& \omega_{k21} &  \cdots &   \omega_{kn1}  \\
 \omega_{k12} & 0&  \cdots&  \omega_{kn2}  \\
\cdots & \cdots& \cdots &\cdots \\
 \omega_{k1n} &  \omega_{k2n}& \cdots &0\\
\end{array}
} \right]
=:G-H. \nonumber
\end{eqnarray*}
Consequently,
\begin{equation*}
G^{-1}= \oplus_{i=1}^n\zeta_{ki}^{-1}
\quad {\rm and}\quad
G^{-1}H= \left[ {%
\begin{array}{cccccccccccccccc}
0& \omega_{k21}\zeta_{k1}^{-1} &  \cdots & \omega_{kn1}\zeta_{k1}^{-1} \\
 \omega_{k12}\zeta_{k2}^{-1}& 0&  \cdots&    \omega_{kn2}\zeta_{k2}^{-1}   \\
\cdots & \cdots& \cdots &\cdots \\
\omega_{k1n}\zeta_{kn}^{-1} & \omega_{k2n}\zeta_{kn}^{-1}     & \cdots &0\\
\end{array}
} \right].
\end{equation*}
Moreover,
$0<\sum_{j=1}^n ( G^{-1}H)_{ij}<1,$ for all $i$.
Hence, $\rho(G^{-1}H)<1$, i.e., $G^{-1}H$  is convergent (see \cite{Plemmons1977}). Obviously, $G^{-1} \geqslant 0$, and  $G^{-1}H \geqslant 0$.  By Theorem $1$ in \cite{Plemmons1977}, $M_k$ is an M-matrix and $M_k^{-1} \geqslant 0$. By the same argument, $X_k$ is an M-matrix and $X_k^{-1} \geqslant 0$. This finishes the proof.

\subsection*{ Network-based RVF meta- population model }

{\it Aedes   population model}
\begin{align}
  \frac{\dif P_{1i}}{\dif t} &= b_{1i} \left(N_{1i}-q_{1i}I_{1i} \right) -\theta_{1i}P_{1i} \label{equation:P1}\\
  \frac{\dif Q_{1i}}{\dif t} &= b_{1i} q_{1i}I_{1i} -\theta_{1i}Q_{1i} \\
 \frac{\dif S_{1i}}{\dif t} &=\theta_{1i}P_{1i}-d_{1i}S_{1i}N_{1i}/K_1-\beta_{21i}S_{1i}I_{2i}/N_{2i}\\
  \frac{\dif E_{1i}}{\dif t} &=\beta_{21i}S_{1i}I_{2i}/N_{2i}-\varepsilon_{1i}E_{1i}-d_{1i}E_{1i}N_{1i}/K_1\\
 \frac{\dif I_{1i}}{\dif t} &=\theta_{1i}Q_{1i}+\varepsilon_{1i}E_{1i}-d_{1i}I_{1i}N_{1i}/K_{1}\\
  \frac{\dif N_{1i}}{\dif t} &=\theta_{1i}(P_{1i}+Q_{1i})-d_{1i}N_{1i}N_{1i}/K_1
\end{align}
\allowdisplaybreaks
{\it Culex  population model}
\begin{align}
  \frac{\dif P_{3i}}{\dif t} &=b_{3i}N_{3i}-\theta_{3i}P_{3i}\\
 \frac{\dif S_{3i}}{\dif t} &=\theta_{3i}P_{3i}-\beta_{23i}S_{3i}I_{2i}/N_{2i}-d_{3i}S_{3i}N_{3i}/K_3\\
  \frac{\dif E_{3i}}{\dif t} &=\beta_{23i}S_{3i}I_{2i}/N_{2i}-\varepsilon_{3i}E_{3i}-d_{3i}E_{3i}N_{3i}/K_3\\
  \frac{\dif I_{3i}}{\dif t} &=\varepsilon_{3i}E_{3i}-d_{3i}I_{3i}N_{3i}/K_3\\
  \frac{\dif N_{3i}}{\dif t} &=\theta_{3i}P_{3i}-d_{3i}N_{3i}N_{3i}/K_3
\end{align}
\allowdisplaybreaks
 {\it Livestock population model}
\begin{align}
  \frac{\dif S_{2i}}{\dif t} &=r_{2i}-\beta_{12i}S_{2i}I_{1i}/N_{1i}-\beta_{32i}S_{2i}I_{3i}/N_{3i} -d_{2i}S_{2i}+\sum^n_{j=1, j \neq
i}\omega_{2ji}S_{2j}-\sum^n_{j=1, j \neq
i}\omega_{2ij}S_{2i}\\
  \frac{\dif E_{2i}}{\dif t} &=\beta_{12i}S_{2i}I_{1i}/N_{1i}
+\beta_{32i}S_{2i}I_{3i}/N_{3i}-\varepsilon_{2i}E_{2i}-d_{2i}E_{2i}+\sum^n_{j=1, j \neq i}\omega_{2ji}E_{2j}-\sum^n_{j=1, j \neq i}\omega_{2ij}E_{2i}\\
  \frac{\dif I_{2i}}{\dif t} &=\varepsilon_{2i}E_{2i}-\gamma_{2i}I_{2i}-\mu_{2i}I_{2i}-d_{2i}I_{2i}+\sum^n_{j=1, j \neq i}\omega_{2ji}I_{2j}-\sum^n_{j=1, j \neq i}\omega_{2ij}I_{2i}\\
\frac{\dif R_{2i}}{\dif t} &=\gamma_{2i}I_{2i}-d_{2i}R_{2i}+\sum^n_{j=1, j \neq i}\omega_{2ji}R_{2j}-\sum^n_{j=1, j \neq i}\omega_{2ij}R_{2i}\\
  \frac{\dif N_{2i}}{\dif t} &=r_{2i}-\mu_{2i}I_{2i}-d_{2i}N_{2i}+\sum^n_{j=1, j \neq
i}\omega_{2ji}N_{2j}-\sum^n_{j=1, j \neq
i}\omega_{2ij}N_{2i}
\end{align}
\allowdisplaybreaks
{\it Human population model}
\begin{align}
  \frac{\dif S_{4i}}{\dif t} &=b_{4i}N_{4i}
-\beta_{14i}S_{4i}I_{1i}/N_{1i}-\beta_{24i}S_{4i}I_{2i}/N_{2i}-
\beta_{34i}S_{4i}I_{3i}/N_{3i}-d_{4i}S_{4i}N_{4i}/K_4\\
 \frac{\dif E_{4i}}{\dif t} &=\beta_{14i}S_{4i}I_{1i}/N_{1i}+\beta_{24i}S_{4i}I_{2i}/N_{2i}+
\beta_{34i}S_{4i}I_{3i}/N_{3i}-\varepsilon_{4i}E_{4i}-d_{4i}E_{4i}N_{4i}/K_4\\
  \frac{\dif I_{4i}}{\dif t} &=\varepsilon_{4i}E_{4i}-\gamma_{4i}I_{4i}-\mu_{4i}I_{4i}-d_{4i}I_{4i}N_{4i}/K_4\\
 \frac{\dif R_{4i}}{\dif t} &=\gamma_{4i}I_{4i}-d_{4i}R_{4i}N_{4i}/K_4\\
  \frac{\dif N_{4i}}{\dif t}&=b_{4i}N_{4i}-\mu_{4i}I_{4i}-d_{4i}N_{4i}N_{4i}/K_4  \label{equation:N4}
\end{align}
\begin{table}[!ht]
\caption{Parameters in the model omitting the node index}
\centering
\begin{tabular}{|p{50pt}p{170pt}  p{75pt}p{25pt} p{120pt}|}
\hline
\textbf{Parameter} & \textbf{Description}& \textbf{Range or value} & \textbf{Units} & \textbf{Source}\\
\hline
$\beta_{12}$ & contact rate: \it Aedes \rm to livestock   & $(0.0021, 0.2762)$&$1/$day &\cite{Canyon1999, Hayes1973, Jones1985, Magnarelli1977, PrattMoore1993, Turell1988, Turell1988b}\\
$\beta_{21}$ & contact rate: livestock to \it Aedes \rm& $(0.0021, 0.2429)$ &$1/$day &\cite{Canyon1999, Hayes1973, Jones1985, Magnarelli1977, PrattMoore1993, Turell1987}\\
$\beta_{23}$ & contact rate: livestock to \it Culex \rm  &$(0.0000, 0.3200)$&$1/$day &\cite{Hayes1973, Jones1985, Magnarelli1977, PrattMoore1993, Turell1987, Wekes1997}\\
$\beta_{32}$ &contact rate: \it Culex \rm to livestock&$ (0.0000, 0.096 )$&$1/$day &\cite{Hayes1973, Jones1985, Magnarelli1977, PrattMoore1993, Wekes1997}\\
$\beta_{14}$ & contact rate: \it Aedes \rm to humans  & &$1/$day & \\
$\beta_{24}$& contact rate: livestock to humans& &$1/$day &\\
$\beta_{34}$& contact rate: \it Culex \rm to humans & &$1/$day &\\
$1/\gamma_2$& recover rate in livestock &$(2, 5) $&days  &\cite{Erasmus1981}\\
$1/\gamma_4$ &recover period in humans &$(4, 7)$ &days  & \cite{Mpeshe2011}\\
$1/d_1$ & longevity of \it Aedes \rm mosquitoes  &$ (3, 60)$ &days &\cite{Bates1970, Moore1993, PrattMoore1993}\\
$1/d_2$ & longevity of livestock  &$(360, 3600)$&days &\cite{Radostits2001}\\
$1/d_3$ & longevity of \it Culex \rm mosquitoes &$(3, 60)$ &days &  \cite{Bates1970, Moore1993, PrattMoore1993}\\
$1/d_4$ & longevity of humans &$ $& days & \\
$b_1$ & egg laying rate of \it Aedes \rm mosquitoes  &&$1/$day &\cite{Bates1970, Moore1993, PrattMoore1993}\\
$b_3$ & egg laying rate of \it Culex \rm mosquitoes & &$1/$day& \cite{Bates1970, Moore1993, PrattMoore1993}\\
$b_4$ & birth rate of humans && $1/$day & \\
$1/\epsilon_1$ &incubation period in \it Aedes \rm mosquitoes  &$ (4, 8)$ &days &\cite{Turell1988}\\
$1/\epsilon_2$ &incubation period in livestock&$(2, 6) $&days &\cite{Peters1994}\\
$1/\epsilon_3$ &incubation period in \it Culex \rm mosquitoes &$(4, 8) $&days &\cite{Turell1988}\\
$1/\epsilon_4$ &incubation period in humans & $(2, 6) $&days & \cite{ Mpeshe2011}\\
$\mu_2$ &mortality rate in livestock & $(0.025, 0.1)$ & $1/$day &\cite{Erasmus1981, Peters1994}\\
$q_1$ & transovarial transmission rate in \it Aedes \rm mosquitoes & $(0, 0.1)$&$1/$day &\cite{Freier1987}\\
$1/\theta_1$&development period of \it Aedes \rm mosquitoes &$(5, 15) $ &days &\cite{PrattMoore1993}\\
$1/\theta_3$&development period of \it Culex \rm mosquitoes &$(5, 15) $&days &\cite{PrattMoore1993}\\
$K_1$ &carrying capacity of \it Aedes \rm mosquitoes &$10000$ & &\\
$K_3$& carrying capacity of \it Culex \rm mosquitoes &$10000$ & &\\
$K_4$& carrying capacity of humans &$100000$ & &\\
$r_{2i}$& livestock recruitment rate&$1$ & $1/$day &  \cite{Mpeshe2011}\\
$\omega_{2ij}$& livestock movement rate from node $i$ to node $j$ &$(0, \frac{1}{n})$ &$1/$day  &\\
\hline
\end{tabular}
\label{table:parameters}
\end{table}
\end{document}